\documentclass[fleqn,usenatbib]{mnras}

\usepackage{newtxtext,newtxmath}

\usepackage{tikz}

\pgfmathsetmacro{\rvec}{.8}
\pgfmathsetmacro{\thetavec}{30}
\pgfmathsetmacro{\phivec}{60}

\DeclareRobustCommand{\VAN}[3]{#2}
\let\VANthebibliography\thebibliography
\def\thebibliography{\DeclareRobustCommand{\VAN}[3]{##3}\VANthebibliography}

\title[Synthetic HROs in a Collapsing Cloud]{Relative Alignment Between Magnetic Fields and Molecular Gas Structure in Molecular Clouds}

\author[R. Mazzei et al.]{
Renato Mazzei,$^{1}$\thanks{E-mail: mazzei@virginia.edu}
Zhi-Yun Li,$^{1}$
Che-Yu Chen,$^{2}$
Laura Fissel,$^{3}$
Mike Chen,$^{3}$
and James Park$^{3}$ 
\\
$^{1}$Department of Astronomy, University of Virginia, 530 McCormick Rd., Charlottesville, Virginia 22904, USA\\
$^{2}$Lawrence Livermore National Laboratory\\
$^{3}$Department of Physics, Engineering Physics, and Astronomy, Queen's University, Kingston, Ontario, Canada
}

\date{Accepted 2023 March 7. Received 2023 March 7; in original form 2023 January 9}

\pubyear{2022}

\begin{document}
\label{firstpage}
\pagerange{\pageref{firstpage}--\pageref{lastpage}}
\maketitle

\begin{abstract}
We compare the structure of synthetic dust polarization with synthetic molecular line emission from radiative transfer calculations using a 3-dimensional, turbulent collapsing-cloud magnetohydrodynamics 
simulation.
The histogram of relative orientations (HRO) technique and the projected Rayleigh statistic (PRS) are considered.
In our trans-Alfv\'enic (more strongly magnetized) simulation, there is a transition to perpendicular alignment at densities above $\sim$$4 \times 10^{3}$ cm$^{-3}$.
This transition is recovered in most of our synthetic observations of optically thin molecular tracers, however for $^{12}$CO it does not occur and the PRS remains in parallel alignment across the whole observer-space.  
We calculate the physical depth of the optical depth $\tau = 1$ surface and find that for $^{12}$CO it is largely located in front of the cloud midplane, suggesting that $^{12}$CO is too optically thick and instead mainly probes low volume density gas.  
In our super-Alfv\'enic simulation, the magnetic field becomes significantly more tangled, and all observed tracers tend toward no preference for perpendicular or parallel alignment.
An observable difference in alignment between optically thin and optically thick tracers may
indicate the presence of a dynamically important magnetic field, though there is some degeneracy with viewing angle. 
We convolve our data with a Gaussian beam and compare it with HRO results of the Vela C molecular cloud.
We find good agreement between these results and our sub-Alfv\'enic simulations when viewed with the magnetic field in the plane-of-the-sky (especially when sensitivity limitations are considered), though the observations are also consistent with an intermediately inclined magnetic field.  

\end{abstract}




\section{Introduction}
All known low-mass star formation occurs in self-gravitating cores and filaments within molecular clouds.
These relatively dense features are formed from diffuse interstellar medium (ISM) gas,
producing local conditions that can facilitate the formation of molecular gas and eventually (in the densest regions, $n \gtrsim 10^{5-6}$ cm$^{-3}$) runaway gravitational collapse that leads to the birth of stars \citep{shu1987,mckee2007}.
The morphology of a molecular cloud (and the geometry of its sub-features, e.g. cores and filaments) is also influenced by the turbulent motions in the gas and the strength and relative orientation of the magnetic field \citep{crutcher2012}. 
The magnetic field not only restricts the flow of gas through tension and pressure forces leveraged across many scales of the cloud, but also can provide direct opposition to gravitational collapse \citep{mestel1956,mouschovias1976}.
One of the main aims in the field of star formation theory, particularly beyond the core scale ($\gtrsim 0.1$ pc), is to develop a general theoretical understanding of the interplay between gas structure, turbulence, and magnetic fields so we may ascertain the dynamical importance of the magnetic field in regulating gas flow through the early and intermediate stages of the star formation process during the molecular cloud's evolution.
A detailed understanding of the 3D magnetic field is required for this task.

Over the last several decades, theoretical studies have used magnetohydrodymaic (MHD) simulations to develop a more comprehensive understanding of the role magnetic fields play in generating the variety of star formation outcomes observed in nature.
Simulations have shown that turbulence at all scales can affect the gas dynamics by forming shearing and converging flows \citep{ostriker2001,nakamura2008}, with turbulence especially having the capability to produce compression that can lead to localized collapse within a cloud \citep{maclow2004}.
This in turn leads to fragmentation that yields overdense gas regions with conditions directly conducive to the formation of stars \citep{scalo1985,ballesteros2007}.
Additionally, in the ISM the magnetic energy density is expected to be approximately in equipartition with the turbulent and gravitational energy densities \citep{heiles2005}.
The magnetic field vector $\boldsymbol{B}$ has some direction, so a large-scale magnetic field introduces an asymmetrical effect on turbulence-driven gas collapse within the cloud.
Particularly, magnetic pressure forces suppress the condensation of gas flows that propagate perpendicular to $\boldsymbol{B}$ \citep{field1965}, allowing shock waves to only flow parallel to the local magnetic field.
As a result, elongated dense gas filaments tend to preferentially form with their crests oriented orthogonal to the mean magnetic field \citep{hennebelle2000,hartmann2001,inoue2007,kortgen2015}. 
\citet{soler2017b} calculated the gradient of the volume density, $\nabla \rho$, and compared it to direction of the magnetic field, $\boldsymbol{B}$, in a turbulent, self-gravitating MHD simulation.
They found that $\nabla \rho$ and $\boldsymbol{B}$ tended to be perpendicular at low density and parallel at high density, with the value of the transition column density dependent on the strength of the magnetic field.
\citet{chen2016} observed a similar result for their analysis of cloud-scale colliding gas flows; overall, the magnetically dominated (sub-Alfv\'enic) post-shock region showed a preference for parallel alignment between gas structures and the magnetic field, but in the densest sub-regions (where there is a transition to super-Alfv\'enic conditions) there was a flip to perpendicular alignment.
Notably, in the more diffuse regions of these simulations gas flows along magnetic field lines were also present, contributing to the parallel alignment in lower density regions. 
These "striation" features have also been seen in observations \citep{andre2014}.

Given the important interplay between magnetic fields and gas structure, magnetic fields are of key observational interest. 
In principle, $\boldsymbol{B}$-field information may be accessed via Stokes $V$ observations, since in the presence of a magnetic field the Zeeman Effect splits the energy levels of some molecules (e.g., CN, OH) into higher- and lower-energy circularly polarized components.
The degree of the splitting is proportional to the strength of the line-of-sight magnetic field, so this effect offers a direct probe of the magnetic field.  
This technique has been successfully used to study the magnetization of dense cores \citep{falgarone2008,troland2008,heiles2004}.
On the cloud-scale, however, where $B_{\rm LOS}$ is expected to be relatively small ($\approx$10 $\mu$G or less), the circular polarization from the Zeeman effect is difficult to detect \citep[][see, however \citet{ching2022}]{goodman1989,crutcher1999}.
In this context, the most effective tool for accessing magnetic field structure is far-infrared and sub-mm linear polarization observations; the dominant source of linear polarization in the large-scale diffuse regions of molecular clouds is "radiative torque" alignment, through which rapidly spinning, effectively oblate dust grains become preferentially oriented with their short axis along the local magnetic field \citep{lazarian2007,hoang2009}.
This yields dust emission that is polarized perpendicular to the magnetic field \citep{davis1951}.

High-resolution dust polarization data provide a means to compute a 2D map of the line-of-sight integrated plane-of-the-sky component of the magnetic field for a given observation.
This technique has been used perhaps most notably by \textit{Planck} to produce 353 GHz all-sky polarization maps at $\sim$10' resolution \citep{planckXIX}.
Additionally, the Balloon-born Large Aperture Submillimeter Telescope for Polarimetry (BLASTPol) has performed high-resolution observations of the Vela C giant molecular cloud ($d \approx 950$ pc).
Vela C is a massive ($\sim$5 $\times$ $10^4$ M$_{\odot}$), relatively cold ($\approx$10-20 K) cloud thought to be in an early phase of its evolution.
Thus relatively unaffected by feedback from massive star formation, it is a pristine laboratory for studying the role of magnetic fields across the many scales of star formation within a cloud.
During its 2012 December run, BLASTPol simultaneously observed Vela C in 250 $\mu$m, 350 $\mu$m, and 500 $\mu$m for a total of 54 hours \citep{galitzki2014}.
In the 500 $\mu$m band, these observations (with correction for ISM dust along the line-of-sight) yielded a 2.5 arcminute ($\sim$0.5 pc, at Vela C distance) resolution map of inferred plane-of-the-sky magnetic field vectors.
Future observing runs by the next generation BLAST Observatory project promise to provide even higher resolution dust polarization observations of several molecular clouds in the Southern Sky.

\citet{soler2013} introduced the histogram of relative orientations (HRO) method for synthetic observations of 3D MHD simulations.
By computing the angle between the local direction of polarization vectors with the gradient of the column density, the HRO provides a way to compare the orientation of the magnetic field with the orientation of gas structures in a 2-dimensional observer space. 
Using this tool, analysis of a subset of \textit{Planck} data taken from ten nearby ($d < 450$ pc) molecular clouds revealed a transition from mostly parallel alignment between the magnetic field field and dense gas structures to mostly perpendicular alignment at $\log{\big(N_{\rm H}/\rm cm ^{-2} \big)} \gtrsim 21.7$ \citep{planckXXXV}.
The HRO technique has also been applied to Vela C, wherein \citet{soler2017} compared BLASTPol polarization data with \textit{Herschel}-inferred column densities.
Their results showed a preference for iso-$N_{\rm H}$ contours to be aligned parallel with the plane-of-the-sky magnetic field along low $N_{\rm H}$ sightlines, and perpendicularly aligned along high $N_{\rm H}$ sightlines. 
\citet{fissel2019} performed a similar analysis, however rather than using column density data, the BLASTPol-inferred magnetic field was instead compared to integrated (Moment 0) line-intensity maps of nine molecular transitions observed with the Mopra telescope.
The advantage of this methodology is that different molecules have different radiative transfer and opacity properties, so their emission may probe different layers of the cloud.
Indeed, it was found that gas structures traced by some molecules (e.g., $^{12}$CO, $^{13}$CO) were preferentially aligned parallel to the magnetic field, whereas higher density tracers (such as C$^{18}$O, CS, and NH$_3$) showed perpendicular alignment.
From these results, combined with simple radiative transfer modeling, they concluded that in Vela C the transition from parallel to perpendicular alignment occurs at $\sim$10$^3$ cm$^{-3}$, between the densities traced by the $J \rightarrow 1 - 0$ transitions of $^{13}$CO and C$^{18}$O.

These observational results suggest a need for more detailed modeling to drive physical interpretation of the gas structures probed by each molecular tracer.
In this work, we perform synthetic molecular line radiative transfer and dust polarization observations of MHD simulations of a turbulent, collapsing molecular cloud threaded by a magnetic field.
We then apply the HRO technique to compare the inferred molecular gas structure from a variety of molecular tracers to the magnetic field information.
Over the course of our analysis, we test a variety of simulation setups to explore how changing the magnetic field strength, viewing geometry, and stage of cloud evolution affect the HRO outcomes.

This paper is organized as follows. 
In Sections \ref{sec:sims}, \ref{sec:polarimetry}, and \ref{sec:lines}, respectively, we introduce our MHD simulation setup, synthetic polarimetry methods, and synthetic line radiative transfer methods.
The details of our HRO calculations are discussed in Section \ref{sec:hrometh}.
We present our main set of results in Section \ref{sec:results}.
This is followed by a discussion in Section \ref{sec:discussion}, that in part focuses on an exploration of optical depth effects for a selection of molecular line observations. By calculating the location of the $\tau = 1$ surface, we develop intuition for which parts of the cloud are being traced by each molecule, thereby linking our observational results back to the underlying 3D physical environment.
In Section \ref{ssec:blastComp} we compare a sub-set of our results with the Vela C results derived from the BLASTPol and Mopra observations.
We perform a beam convolution on these selected synthetic data to facilitate a more direct comparison.
The main conclusions are summarized in Section \ref{sec:conc}.

\section{Numerical Simulations} \label{sec:sims}

The simulations in this work were performed using ATHENA, a 3-dimensional grid-based MHD code \citep{stone2008}.
Our simulations are in full 3D with ideal MHD assumptions.
Our simulation models a turbulent, initially spherical ball of dense gas, embedded in a low-density ambient environment threaded by a uniform magnetic field.
This is intended to mimic a typical isolated molecular cloud, collapsing under the influence of gravity against turbulent and magnetic support.
We prescribe an isothermal equation-of-state, with temperature $T = 10$ K.

Our cloud is initialized as a pseudo-Bonner Ebert sphere with number density

\begin{equation}
    n(r) = \frac{n_0}{1+(r/r_c)^2}\,.
    \label{eq:bonnerSphere}
\end{equation}

We set the central density $n_0 = 2000$ $H_2$/cm$^3$ and choose $r_c = 0.5R$, where $R = 2$ pc is the radius of the cloud. Outside this radius the density has a sharp cutoff, linearly decreasing from $n(R)$ to the ambient density $n(R)/100$ over a shell of width $dr = 0.01R$.
The cloud is placed in a box with side length $L = 5$ pc on a $256 \times 256 \times 256$ fixed grid. 
Therefore, each simulation cell has a width of $\sim$0.02 pc.
We adopt outflow boundary conditions.

Initial gas velocities are set by prescribing a perturbation for each cell, sampling from a Gaussian random distribution with power-law turbulence $v_k \propto k^{-2}$ and the amplitude of the velocity perturbation set to $\sigma_v = 10 c_s$. The turbulence is not driven. 

Finally, the initial magnetic field is set as $\boldsymbol{B_0} = (0,0,B_z)$ where $B_z$ is parameterized by the Alfven Mach number $M_A$:

\begin{equation}
    M_A = \frac{\sigma_v \sqrt{4\pi\rho_0}}{B_z}\,.
    \label{eq:cloudAlfvenMach}
\end{equation}

Here we adopt the conventional assumption that $n_{\rm He} = 0.1 n_{\rm H}$, such that $\rho_0 = 2.8 m_{\rm p} n_0$.
We produce two simulation versions, one with a weaker (super-Alfv\'enic) initial magnetic field ($B_{z,0} = 16$ $\mu$G; hereafter called Model W), and one with a stronger (trans-Alfv\'enic) initial magnetic field ($B_{z,0} = 58$ $\mu$G; hereafter called Model S).
These values correspond to $M_A = 4$ and $M_A = 1$ for Model W and Model S, respectively.

To investigate how our results change through different stages of the cloud's evolution, in this work we examine synthetic data collected from uniformly sampled snapshots of these two models, taken between 0.25 Myr and 1.25 Myr after the simulation start time. 
To probe different viewing geometries, we also select a variety of angles at which to observe the cloud.
The orientation of an observer's line-of-sight $s$ relative to simulation coordinate system can be parameterized with two parameters: its inclination $i$ away from the $z$-axis, and its position angle $\text{PA}$ away from the $x$-axis in the $xy$-plane.
Since our simulations do not have any preference in azimuthal direction around the $z$-axis (i.e, the direction of $\boldsymbol{B_0}$), we can effectively sample the unique geometries of the observer space by adjusting only $i$ and setting $\text{PA}=0$.
Hereafter, we will sometimes refer to the view with $i=0^{\circ}$ as the $B_{0,\rm LOS}$ view (since from this view, the direction of the initial magnetic field is along the line-of-sight), and the view with $i=90^{\circ}$ as the $B_{0,\rm POS}$ (since from this view, the direction of the initial magnetic field is in the plane-of-the-sky).

\begin{figure*}
\centering
\includegraphics[width=0.93\textwidth]{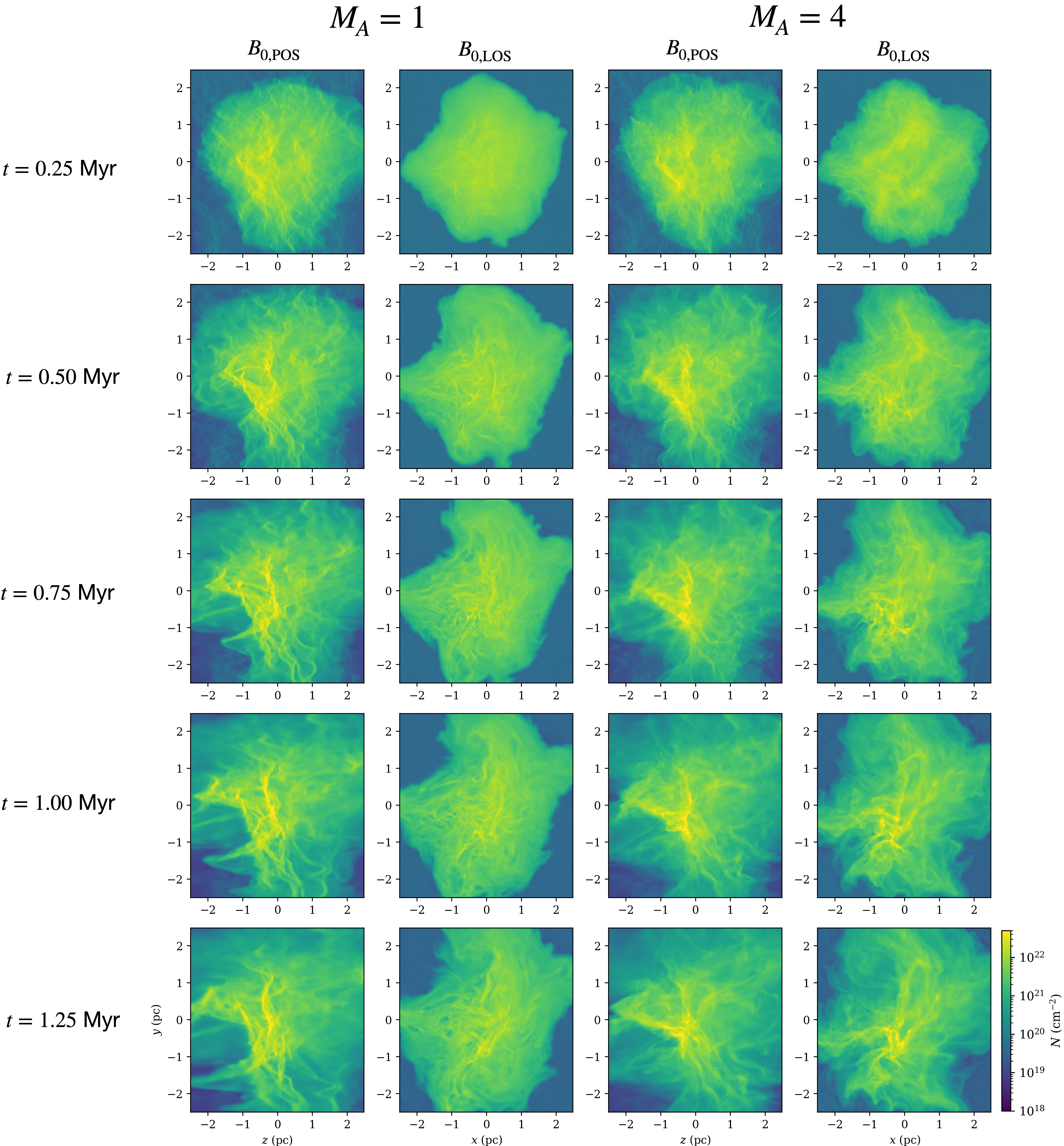}
\caption{Column density maps for our strong and weak magnetic field simulations ($M_A = 1$ and $M_A = 4$), viewed with the initial magnetic field in the plane-of-the-sky ($B_{\rm 0,POS}$, first and third column) and along the line of sight ($B_{\rm 0,LOS}$, second and fourth column), as a function of time elapsed after simulation initialization.} 
\label{fig:coldens}
\end{figure*}

Presented in Figure \ref{fig:coldens} are column density maps for our simulations, as observed from the $B_{0,\rm POS}$ view and the $B_{0,\rm LOS}$ view at the five time steps we will consider in this work.
These plots demonstrate the important effect that a strong magnetic field strength can have on gas dynamics within a cloud, and moreover how the orientation of the $\boldsymbol{B}$-field relative to the observer can impact the apparent gas structure even for an identical physical environment.
Furthermore, the two magnetic field strengths produce different temporal evolution, which is especially evident at later times.
For example, at the $t = 1.25$ Myr time step the two viewing geometries for the stronger field $M_A = 1$ simulation are clearly distinct, with the $B_{0,\rm POS}$ view showing filamentary structure orthogonal to the magnetic field.
Meanwhile, the two views of the $M_A = 4$ simulation are similar. 
This isotropic structure formation suggests a gravitationally dominated collapse scenario in which the magnetic field has less dynamical importance.

\begin{figure*}
\centering
\includegraphics[width=0.97\textwidth]{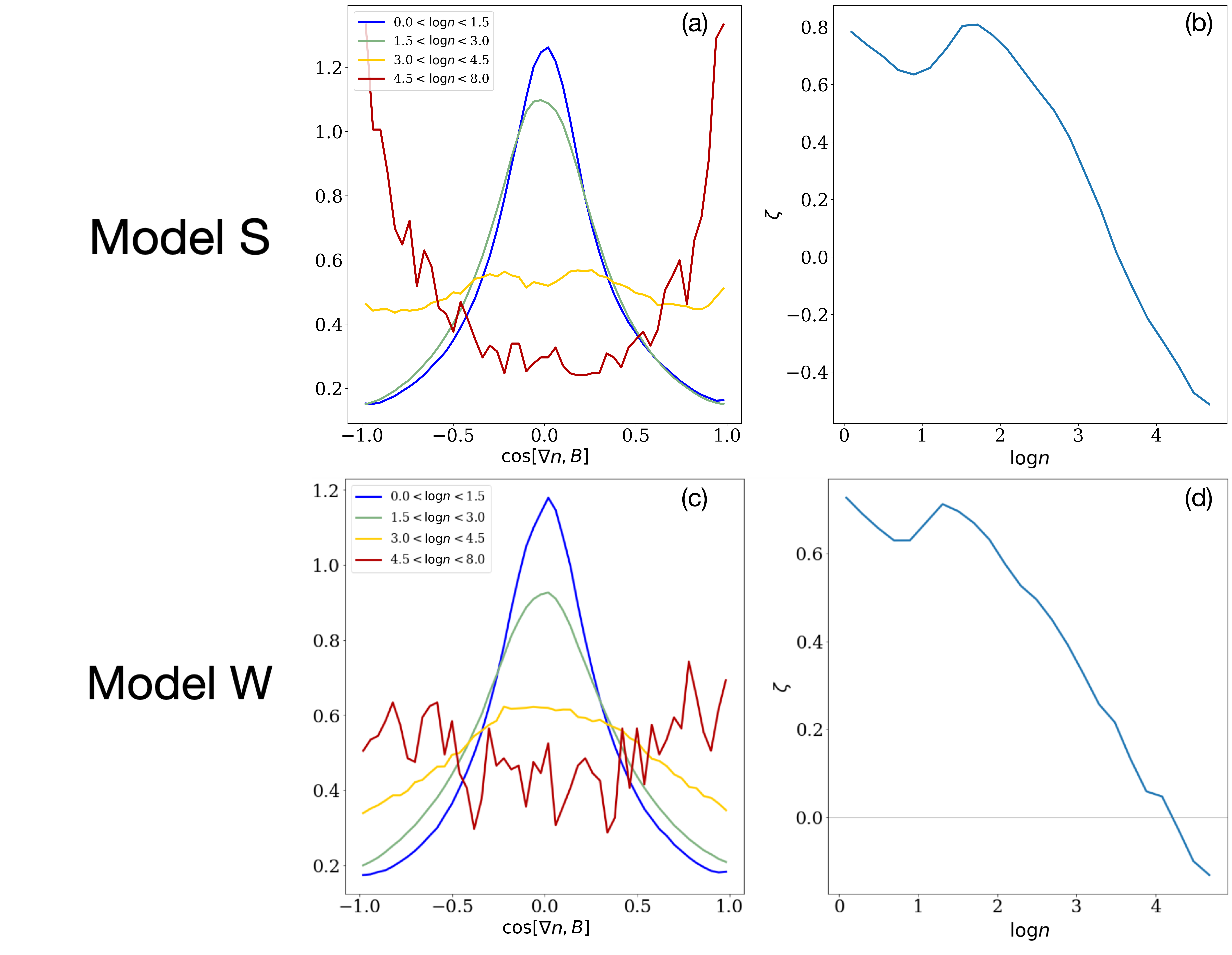}
\caption{Plots depicting the 3D orientation of the magnetic field relative to density structures in our Model S (Panels (a) and (b)) and Model W (Panels (c) and (d)) simulations at the $t = 0.75$ Myr time step. \textbf{Panel (a)}: The average angle between $\nabla n$ and $\boldsymbol{B}$ for a few density bins. At low density there is a local 3D preference for parallel alignment between the magnetic field and gas structures (indicated by a value of $\cos{[\nabla n, B]} = 0$) . This flips to a perpendicular preference as the density is increased. \textbf{Panel (b)}: $\zeta$ as a function of $\rho$. 
In this particular frame, the crossover to perpendicular alignment occurs when $n \gtrsim 4 \times 10^{3}$ cm$^{-3}$. \textbf{Panels (c) and (d)}: Same plots for Model W, the weaker magnetic field case. There is still a preference for parallel alignment in the low density regions, but the highest density regions no longer show much preference for perpendicular alignment.} 
\label{fig:cheyuplots}
\end{figure*}

Also of interest is the relationship between the 3D orientation of the magnetic field and density structures.
Figure \ref{fig:cheyuplots} quantifies this relative orientation for the $t = 0.75$ Myr time step in Model S and Model W.
In this 3D view, the magnetic field tends to be parallel to the orientation of local gas structures when the density is low, but the orientation begins to flip at higher densities. 
This effect is especially clear in Model S, wherein (due to the stronger magnetic guiding gas structure formation) there is a clear preference for perpendicular alignment at the highest densities.
The transition threshold occurs at  $n \gtrsim 4 \times 10^{3}$ cm$^{-3}$.
Meanwhile, in Model W there is significantly less alignment preference at the highest densities.
We quantify this using the HRO shape parameter $\zeta$.
Positive $\zeta$ corresponds to a preference for parallel alignment, and negative $\zeta$ corresponds to a preference for perpendicular alignment \citep[see][]{planckXXXV,chen2016}.

\section{Synthetic Polarimetry} \label{sec:polarimetry}
To perform mock polarimetry on these data, we follow the literature-standard practice for the computation of synthetic Stokes parameters \citep[e.g.,][]{planckXX,chen2016}, writing the following expressions for $I$, $Q$, and $U$:




\begin{align}
    I &= N-p_0 N_2 \,, \\
    Q &= p_0 \tilde{Q}\,, \\
    U &= p_0 \tilde{U}\,,
\end{align}
where 
\begin{align}
N &= \int n ds \,, \label{eq:cd} \\
N_2 &= \int n \Big(\frac{B_x^2+B_y^2}{B^2}-\frac{2}{3}\Big)ds\,, \\
\tilde{Q} &= \int n \Big(\frac{B_y^2-B_x^2}{B^2}\Big)ds \,, \\
\tilde{U} &= \int n \Big(\frac{2 B_x B_y}{B^2}\Big)ds \,.
\end{align}
Note that since it is usually the case that $N_2 << N$, the Stokes $I$ generally gives a good approximate probe of column density \citep{king2018}.
For this work, we adopt the literature standard value of $p_0 = 0.15$.

In performing these integrations for each pixel column along the chosen line-of-sight, we produce 2-dimensional maps of $N$, $I$, $Q$, and $U$ in synthetic observer space.
Polarization fraction $p$ and polarization angle $\chi$ (measured in the plane-of-the-sky) for each pixel are then calculated as
\begin{align}
    p = \frac{\sqrt{Q^2+U^2}}{I}
\end{align}
and 
\begin{align}
    \chi = \frac{1}{2} \arctan{(U,Q)}\,.
\end{align}

\begin{figure*}
\includegraphics[width=\textwidth]{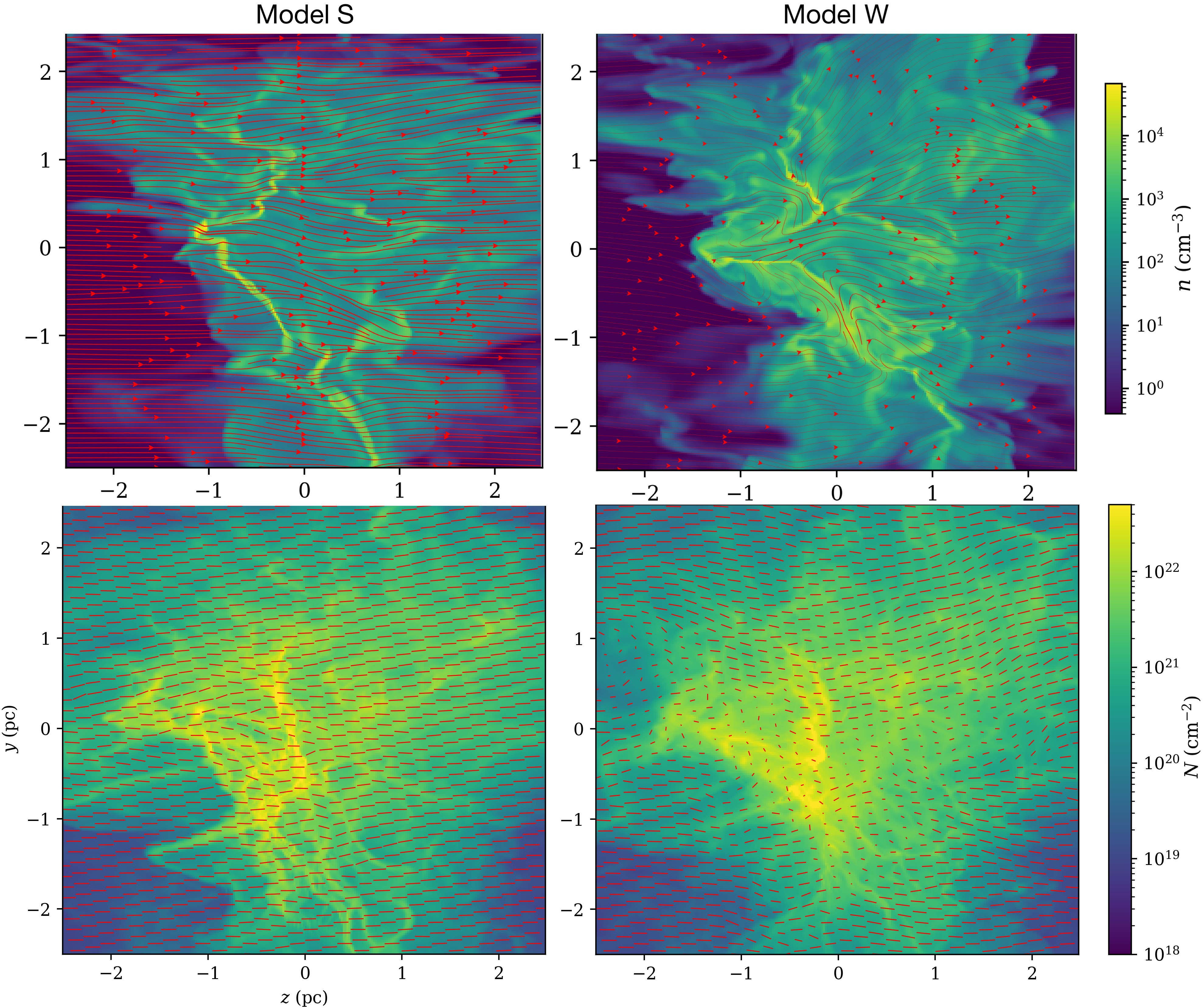}
\caption{\textbf{\textit{Panels (a) and (b):}} Volume density midplane cuts through our simulation space, for Model S and Model W, respectively, from the $t=0.75$ Myr frame. Overplotted on each are streamlines of the magnetic field vectors in the plane. \textbf{\textit{Panels (c) and (d):}} Corresponding polarization vector maps, plotted overtop the column density maps. There is significantly more de-polarization in the high column density regions of the weak field simulation, due to increased field line tangling in those regions.}
\label{fig:streams}
\end{figure*}

Plotted in Figure \ref{fig:streams} are polarization vectors for both the weak and strong magnetic field cases at the $t = 0.75$ Myr time step, as viewed with the line-of-sight along the $x$-axis (the $B_{\rm POS}$ view).
We also provide a comparison with $\boldsymbol{B}$-field "streamlines" taken from the midplane ($x = 0$) of our simulations. 
Just as the magnetic field affects the gas structure, its impact is also evident in the polarization.
In the stronger magnetic field case ($M_A = 1$) the $\boldsymbol{B}$-field has largely maintained its initialized orientation (i.e., $\boldsymbol{B} \approx (0,0,B_z)$), and the polarization vectors reflect this.
By contrast, there is significantly more change of direction of the $\boldsymbol{B}$-field in the $M_A = 4$ simulation.
In kind, the polarization vectors are more disordered.
There are also some regions with significant de-polarization.
Absent any dust grain physics effects (which are not considered here), this is caused by the magnetic field being bent away from the plane-of-the-sky or being "tangled" in such a way that in projection it has less apparent local preferred direction.     

Notably, these vector maps are qualitatively consistent with the 3D alignment preference data depicted in Figure \ref{fig:cheyuplots}.
This demonstrates that the 2D observables have some power in diagnosing the physical conditions underpinning our simulations.
Furthermore, the data are consistent with visual inspection of the volume density slices.
This is made especially clear by panel (a) of Figure \ref{fig:streams} for the strong field case.
The low- and intermediate-density parts of the cloud have structures that appear to flow along the $\boldsymbol{B}$-field lines, whereas the densest filament is orthogonal to the magnetic field.
In contrast, the orthogonal alignment in the densest regions is less clear in the weak field case (see panel (b) of Figure \ref{fig:streams}). 
This contrast provides a way to distinguish the strong and weak field cases through polarimetric observations (see Section \ref{ssec:fieldstrength}).


\section{Synthetic Molecular Line Observations} \label{sec:lines}

To perform our HRO analysis, we require a probe of gas structure.
Molecular line data are a useful tool for this purpose in the cloud environment. 
By observing a variety of molecular tracers (each with distinct excitation conditions and optical depths), we may leverage the information available in these maps in combination with polarization data to learn about the relationship between the (line-of-sight integrated) magnetic field and gas structure at a variety of cloud depths.

\subsection{Line Simulation Methods} \label{ssec:lineMeth}
Line radiative transfer (LRT) simulations were performed using the RADMC-3D\footnote{https://www.ita.uni-heidelberg.de/~dullemond/software/radmc-3d/} radiative transfer code. 
A synthetic LRT observation requires the following physical input data for each cell in the observed 3D space:
\begin{itemize}
    \item Gas number density $n$ 
    \item Gas temperature $T$
    \item Gas velocity $ \boldsymbol{v} = (v_x,v_y,v_z)$
    \item Abundance $X$ of the molecular species being observed
    \item Molecular transition data for the species being observed
\end{itemize}

In addition to $n$, $T$, and $\boldsymbol{v}$, which can be read directly from the simulation, 
molecular transition data were imported from the Leiden Atomic and Molecular DAtabase \citep[LAMDA;][]{schoier2005}.
Abundance is a parameter that may be set freely. In Table \ref{table:fidAbundances}, we list the fiducial values chosen for each molecular species used in this work. Our choices are based on data from a few different molecular cloud observation programs \citep{fuente2019,maret2006,morgan2013}.
The abundance ultimately prescribes the number density of the observed species placed 
in each cell of the simulation:
\begin{align*}
    n_{\rm species}(x,y,z) = X_{\rm species} n(x,y,z) \,.
\end{align*}

After setting up these physical parameters, we then must choose the position of the detector. We place it at a simulated distance $d = 950$ pc away from our molecular cloud (i.e., the approximate \textit{Gaia}-constrained distance to Vela C), and adjust its orientation relative to the initial magnetic field by setting $i$.
We set the source velocity to $v_0 = 0$ km~s$^{-1}$ relative to the observer.


To carry out each LRT calculation $10^5$ unpolarized background photons are initialized behind (relative to the detector) the 3D simulated cloud, and the radiative transfer is iteratively calculated until all photons have been propagated. We find this to be a sufficient number of photons for convergence in our case.  In our main set of models, we assume local thermodynamic equilibrium (LTE) for level population calculations. To assess the effect of this assumption on our results, we also perform (for a sub-set of our parameter space grid) synthetic observations that use the non-LTE large velocity gradient (LVG) approximation.
For these LVG runs, a collisional partner (with corresponding number density) must be input into the simulation. We adopt the standard choices here, using $H_2$ as the collisional partner with $n_{H_2} = n$.

To generate images, photons are ray traced to the $256\times 256$ pixel detector, and the emission is recorded.
We set the detector to observe the specific intensity $I_{\nu}$ at $n_{\rm freq} = 111$ frequencies in range $[v_0-3\text{ km/s},v_0+3\text{ km/s}]$ about the rest frequency of each of our lines, producing data with a velocity resolution of $\Delta \nu = 0.05$ km s$^{-1}$. 

\begin{table*}
\caption{Molecular line transition information and abundance data for each species simulated in this work. Choices of lines to synthetically observe were motivated by Mopra observations from \citet{fissel2019}. \label{table:fidAbundances}}
\begin{tabular}{ccccc}
\hline
Species & Transition & Rest Frequency (GHz) & Fiducial Abundance & Abundance Reference \\
\hline
$^{12}$CO   & $J = 1 \rightarrow 0$     & 115.2712  & $1.4 \times 10^{-4}$ & \citet{fuente2019} \\
$^{13}$CO   & $J = 1 \rightarrow 0$     & 110.2013  & $2.3 \times 10^{-6}$ & $X(^{12}$CO$) / 60$ \\
C$^{18}$O   & $J = 1 \rightarrow 0$     & 109.7822  & $2.3 \times 10^{-7}$ & $X(^{12}$CO$) / 600$ \\
N$_2$H$^+$     & $J = 1 \rightarrow 0$     & 93.1730  & $5.0 \times 10^{-10}$ & \citet{maret2006}\\
HNC         & $J = 1 \rightarrow 0$     & 90.6636  & $5.0 \times 10^{-10}$ & \citet{fuente2019}\\
HCO$^{+}$        & $J = 1 \rightarrow 0$     & 89.1885  & $1.3 \times 10^{-8}$ & \citet{fuente2019}\\
HCN         & $J = 1 \rightarrow 0$ (hfs)     & 88.6319  & $1.0 \times 10^{-9}$ & \citet{fuente2019}\\
CS          & $J = 1 \rightarrow 0$     & 48.9910  & $2.0 \times 10^{-8}$ & \citet{fuente2019}\\
NH$_3$      & (1,1)(hfs)     & 23.6945  & $1.0 \times 10^{-8}$ & \citet{morgan2013}\\
\hline
\end{tabular}
\end{table*}

\subsection{Computation of the Moment 0 Maps} \label{ssec:lineMeth}

For the purposes of the analysis in this work, we are interested in the integrated intensity map 
$I$ of the line emission.
Our simulations provide us with $I_{\nu}$ for each simulated frequency.
Since our molecular lines are in the Rayleigh-Jeans limit, we may convert each of these values to brightness temperatures,
\begin{equation}
    T_B(\nu) = \frac{I_{\nu} c^2}{2 k \nu^2}\,,
\end{equation}
where $c$ is the speed of light and $k$ is the Boltzmann constant.
To compute $I$ (in K km s$^{-1}$ units), we then sum over the observed frequencies :
\begin{equation}
    I = \sum_i^{n_{\rm freq}} T_B(v) \Delta v \,.
\end{equation}
Note that implicit in this summation is a conversion from frequency space to velocity space, where in this regime radial velocities (relative to line rest frequency $\nu_0$) may be computed as
\begin{equation}
    v = \frac{c (\nu_0 - \nu)}{\nu_0}\,.
\end{equation}

\section{Analysis Methods} \label{sec:hrometh}

Dust polarization observations provide us with a proxy for the plane-of-the-sky orientation of the magnetic field, and Moment 0 intensity maps from our molecular tracers give information on the plane-of-the-sky orientation of the distribution molecular gas.
In this work, we use the analysis tools described below to perform pixel-by-pixel comparisons of these two sets of data.

\subsection{The Histogram of Relative Orientations}

As described in \citet{soler2017}, alignment of the magnetic field vector with the direction of an iso-$I$ filament is equivalent to alignment of the electric field vector $\boldsymbol{E}$ with the gradient of the local intensity, $\nabla I$.
The relative orientation angle $\phi$ is calculated as 
\begin{equation}
    \phi = \arctan(|\nabla I \times \hat{E}|, \nabla I \cdot \hat{E})\,,
\end{equation}
with $\phi = 0^{\circ}$ corresponding to (local) parallel plane-of-the-sky alignment between the magnetic field and gas structure, and $\phi = 90^{\circ}$ corresponding to orthogonal alignment. 
To obtain a statistical understanding of the relative alignment across the observer-space, we may plot the Histogram of Relative Orientations (HRO) for all $\phi$ in the map.

\subsection{The Projected Rayleigh Statistic}

In addition to the HRO technique, we can use the Projected Rayleigh Statistic (PRS) to distill the information provided by the relative orientation calculation into a single parameter that characterizes the global alignment across the observer space.
As described in \citet{jow2018}, the PRS $Z_x$ is a metric that indicates whether there is a preference for parallel or perpendicular alignment within a set of independent angle measurements. 
Taking $\theta = 2 \phi$ , such that $\theta = 0$ corresponds to parallel alignment and $\theta = \pi$ corresponds to perpendicular alignment, the PRS $Z_x$ is given as 
\begin{equation}
    Z_x = \frac{\sum_i^{n_{\rm ind}} \cos \theta_i}{\sqrt{n_{\rm ind}/2}}\,.
\end{equation}
In our case, the independent samples are the values of $\theta$ in each pixel across the map, so $n_{\rm ind} = 256^2$.
Positive values of $Z_x$ suggest a tendency toward parallel alignment between the magnetic field and gas filaments, and negative values of $Z_x$ suggest a tendency toward perpendicular alignment.
The larger the value of $|Z_x|$, the greater the alignment preference.

We correct for PRS oversampling using the white noise map protocol described in \citet{fissel2019}. All of our PRS measurements have a 3-sigma uncertainty of $\pm$1.

\section{Results}\label{sec:results}

To assess HRO results for a variety of scenarios, we performed LRT synthetic calculations for several different simulation set-ups by independently adjusting a few key parameters, which we split into "main parameters" ($M_A$, $i$, $t$, and molecular species) and "auxiliary parameters" (listed in Table \ref{table:parameterGrid}).
Presented in this section are the results obtained from adjusting the main parameters.
This "main grid" of models establishes a baseline for understanding how the outcome of our HRO analysis is affected by the intrinsic physics of the cloud and viewing geometry effects.
These computations were performed using the LTE radiative transfer assumption, with abundances set by the values from Table \ref{table:fidAbundances} and no beam convolution applied.
Adjustments to the auxiliary parameters are addressed in Discussion sections \ref{ssec:caseStudy}, \ref{ssec:lvgComp}, and \ref{ssec:blastComp}, respectively. 

To establish a reference point to cross-compare the results gathered from these many parameter adjustments, we define a fiducial frame: the $t=0.75$ Myr snapshot of Model S ($M_A = 1$), as observed in the $B_{0,\rm POS}$ view  ($i=90^{\circ}$).
This frame was chosen because it is a good representative frame for capturing the many elements of our simulation.
It is about halfway through the runtime, at which point clear filaments have started to form and gravitational collapse is beginning to take hold, despite the initial turbulence.
Furthermore, since dust polarimetry captures the plane-of-the-sky component of the magnetic field, this view gives us the best direct representation of the global $\boldsymbol{B}$-field.

Our parameter space exploration is organized as follows.
First, in Section \ref{ssec:molTrace} we investigate how the HRO and PRS results change for different molecules, as observed in the fiducial frame and as a function of time. 
In Section \ref{ssec:fieldstrength} we present the results for adjustments to the magnetic field strength (i.e, Model S vs. Model W), and in Section \ref{ssec:inclination} the effect of inclining $\boldsymbol{B_0}$ between 0$^{\circ}$ and  90$^{\circ}$ relative to the observer.

\begin{table*}
\caption{A summary of the parameter space we explore in our synthetic observations. For most parameters, fiducial values are bolded. In addition to testing two different magnetic field strength simulations at different stages of evolution, we adjust several parameters related to the observed molecular species and viewing geometry.}
\label{table:parameterGrid}
\begin{tabular}{llll}
\hline
& Parameter & Range Explored & Notes \\
\hline
Main Parameters & $M_A$   & \textbf{1}, 4    & Corresponding to $B_{z,0} = 58$ $\mu$G and $B_{z,0} = 14$ $\mu$G, respectively\\
& $t$ (Myr)     & 0.25, 0.5, \textbf{0.75}, 1.0, 1.25 & Simulation time \\
& $i$ ($^{\circ}$) & 0, 30, 60, \textbf{90} & Inclination of $\boldsymbol{B_0}$ relative to the observer \\
& Molecular Tracer & $^{12}$CO, $^{13}$CO, C$^{18}$O, CS, HCO$^{+}$&  \\
& & N$_{2}$H$^+$, HNC, HCN, NH$_{3}$ & \\
& & & \\
Auxiliary Parameters & Molecular Abundance & $10^{-10}$ to $10^{-4}$ & Fiducual values given in Table \ref{table:fidAbundances} \\
& Radiative Transfer Mode & LTE, LVG & Using RADMC3d \\
& Beam Size (pc) & \textbf{0.02,} 0.2, 2.0 & Computed with Gaussian convolution\\
\hline
\end{tabular}
\end{table*}

\subsection{Molecular Tracers}\label{ssec:molTrace}

\begin{figure*}
\centering
\includegraphics[width=0.98\textwidth]{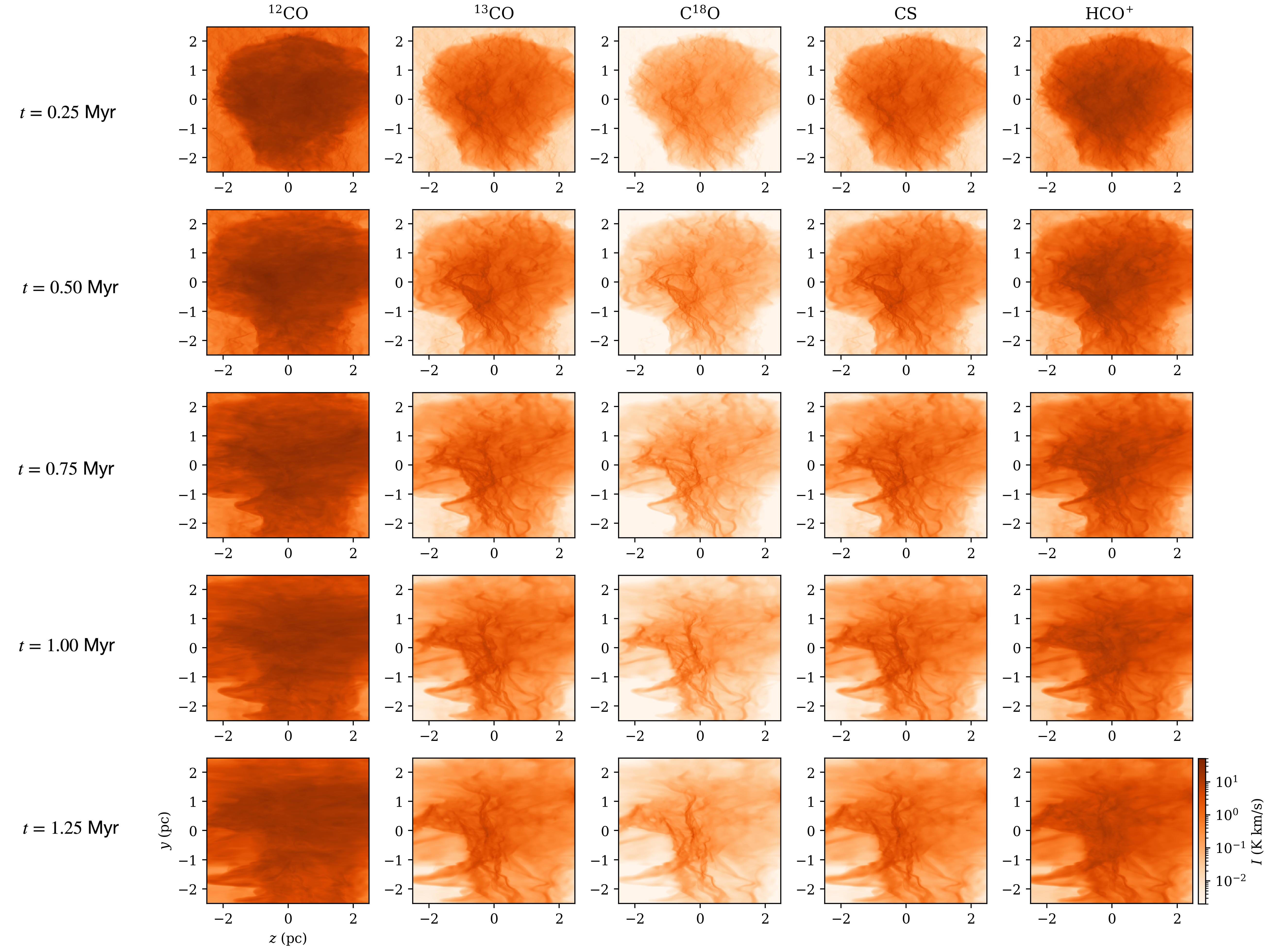}
\caption{Moment 0 Maps for a selection of our modeled tracers, viewed with the magnetic field in the plane-of-the-sky and plotted at the same timesteps shown in Figure \ref{fig:coldens}. The difference in distribution of intensity across the tracers is clear. In $^{12}$CO, for example, it is not possible to see any of the signature of high density filamentary structure that is present in the other CO isotopologues.} \label{fig:fidMoment0}
\end{figure*}

To synergize with the existing observational data, we chose to perform synthetic observations for the nine (ground-state) molecular line transitions studied in \citet{fissel2019} for the Vela C molecular cloud (see Table \ref{table:fidAbundances} for species and rest frequencies).
Moment 0 maps (at fiducial abundance) are presented in Figure \ref{fig:fidMoment0}, plotted at $t = 0.25$, 0.5, 0.75, 1.0, and 1.25 Myr.
The implied gas structure varies across many of the molecules.
The tracer with the highest abundance in the set, $^{12}$CO, yields a map that is significantly more uniform in intensity than the others.
It is optically thick, with $\tau > 1$ across much of the cloud.
Meanwhile, tracers like CS and the isotopologue C$^{18}$O reveal much more varied structure.
These maps almost exactly match the column density calculated simply by summing up the volume density along the line-of-sight. 
They represent the optically thin case, and are more effective at tracking the evolution of high density sub-regions as time progresses (see e.g. \ref{fig:fidMoment0}). 
Whereas the $^{12}$CO map remains relatively uniform in intensity at later times, the CS map (for example) traces the condensation of mass into
filaments.  
There are also some molecules that sit between these extremes with more intermediate $\tau$, like HCO$^{+}$ and $^{13}$CO.

\subsubsection{HROs for Molecular Tracers} \label{sssec:hroresults}

\begin{figure*}
\centering
\includegraphics[width=0.98\textwidth]{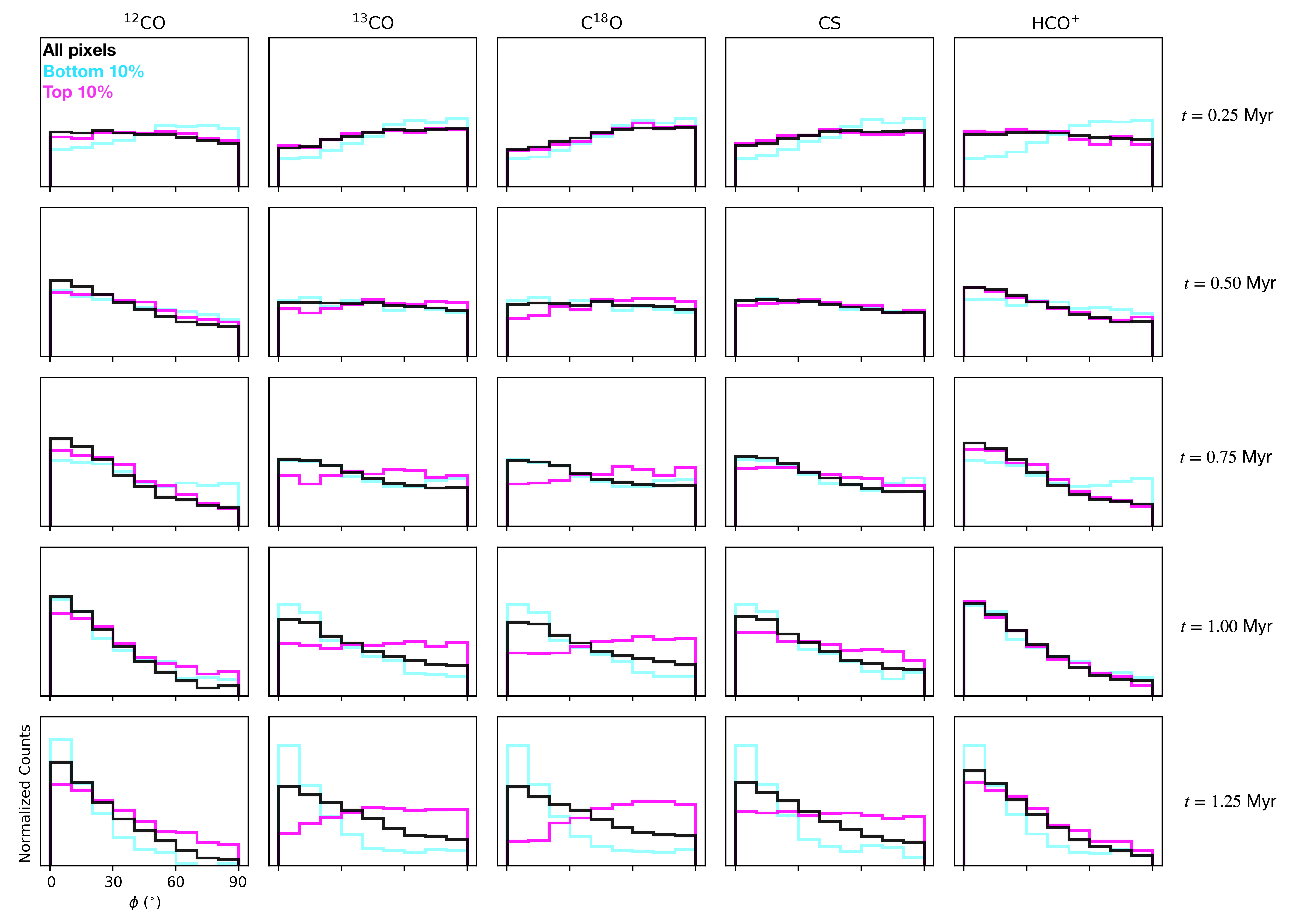}
\caption{Histograms of relative orientation for the same subset of tracers shown in Figure \ref{fig:fidMoment0}. Included are HRO results that include all observer-space cells in the computation (black), the top 10\% of cells in molecular intensity Moment 0 map value (pink), and the bottom 10\% of cells in molecular intensity Moment 0 map value (blue).  As discussed in the text (Section \ref{sssec:hroresults}), several of the tracers generally show a preference for perpendicular alignment (indicated by a positive HRO slope) in the highest intensity cells, as expected for dense filaments threaded by a strong magnetic field (as in panels (a) and (c) of Figure \ref{fig:streams}). The major exception to this is $^{12}$CO, which universally shows a strong preference for parallel alignment.} \label{fig:hros}
\end{figure*}

Shown in Figure \ref{fig:hros} are the HROs computed using the intensity gradients of the maps in Figure \ref{fig:fidMoment0}.
Histograms with negative slopes (i.e., more counts in low $\phi$ bins) indicate a preference for parallel alignment between molecular structure and the magnetic field, and histograms with positive slopes (i.e., more counts in high $\phi$ bins) indicate a preference for perpendicular alignment.
In addition to HROs that incorporate all the pixels in the synthetic observation (plotted in black), we also include versions that only use the 10\% lowest intensity cells and the 10\% highest intensity cells.

Starting with the black ("all cells") curves, we can observe two notable trends.
First, there is in general a clear distinction between $^{12}$CO and all the other tracers.
At early times ($t=0.25$ Myr), $^{12}$CO has a slightly negative slope (i.e., a slight preference for parallel alignment), while the others have a slight positive slope.
At this point in the simulation, the cloud still has relatively little density contrast compared to later in its evolution (excluding the ambient low-density gas outside the main cloud), because it takes time for density structures to fully develop out of the turbulence imposed at the start of the simulation. 
However, some moderately high-density filaments have started to form with their ridges mainly orthogonal to the magnetic field.
As reflected in the intensity maps, the higher density tracers like C$^{18}$O and CS pick up these filaments, producing a modest preference for perpendicular alignment.
Meanwhile, these filaments are not present in the $^{12}$CO emission, and as a result it does not take on the same HRO shape.
Its slight preference for parallel alignment is due to a somewhat subtle effect.
As demonstrated in \citet{xu2019}, the field-aligned filaments in the low-density regions are a natural consequence of the magnetically induced anisotropy in the cloud turbulence. Similar filaments are also found in other simulations, such as \citet{chen2017}.
These "striations" are visible in the $^{12}$CO maps, running parallel to the global $\boldsymbol{B}$-field direction along the $z$-axis.

The second clear trend we see in the "all cells" HROs is that as time advances the tendency toward parallel alignment is enhanced. 
This is present for all molecules observed, and the cause is two-fold.
The main effect is that as time progresses, the mass becomes more concentrated as it is transported into the filaments near cloud center.
As a result the majority of the volume becomes low-density, and alignment preference is dominated by the magnetically-aligned striations.
The secondary effect, which leads to the same result, is that the cloud expands over time. 
This, again, results in more of the volume of the simulation box being occupied by low density gas flowing toward the filaments.

The global HRO shape is largely determined by what fraction of the synthetic observation is occupied by low-density striations vs. higher-density filaments.
As the pink histograms in Figure \ref{fig:hros} show, however, a different trend is revealed when we only consider the top 10\% of cells by intensity.
In this case, the high density tracers show HROs with more perpendicular alignment as time progresses.
Effectively, with this emission cutoff it is possible to capture the perpendicular alignment in the filaments without having the lower-density regions dilute the overall HRO result.
Meanwhile, the HRO slope for $^{12}$CO remains about the same as for the "all cells" case.
This is because the $^{12}$CO is too optically thick to probe the higher density filaments (see further discussion in Section \ref{ssec:depthDisc}).

\subsubsection{PRS Results for Molecular Tracers}

HROs are useful for visualizing alignment preference across the observer space.
The PRS distills this down to a single number, giving us a sense of global alignment between the magnetic field and molecular structure in a particular snapshot.
To understand how this quantity changes as the cloud evolves, in panels (a) and (b) Figure \ref{fig:prs} we plot $Z_x$ as a function of time for each of our tracers (using the fiducial model), for the cases where we use all cells and just the top 10\% of cells in the computations, respectively.
This plot clearly illustrates some of the general trends discussed in the previous section. 
In the "all cells" plot, the PRS systemically increases for all molecules, and at later times (when gravitational infall begins to dominate) $Z_x$ is large ($>$10) for all tracers.
Meanwhile, in the high-intensity version the high density tracers remain with a negative PRS throughout the cloud evolution.
This, again, is a reflection of the perpendicular alignment preference that persists in the cloud's high-density filaments. 
Note that, compared to all the other tracers besides $^{12}$CO, HCO$^{+}$ has a relatively high PRS.
This is because of the selected abundance, a point we discuss in more detail in Section \ref{ssec:caseStudy}.

\subsection{Magnetic Field Strength}\label{ssec:fieldstrength}

\begin{figure*}
\centering
\includegraphics[width=0.98\textwidth]{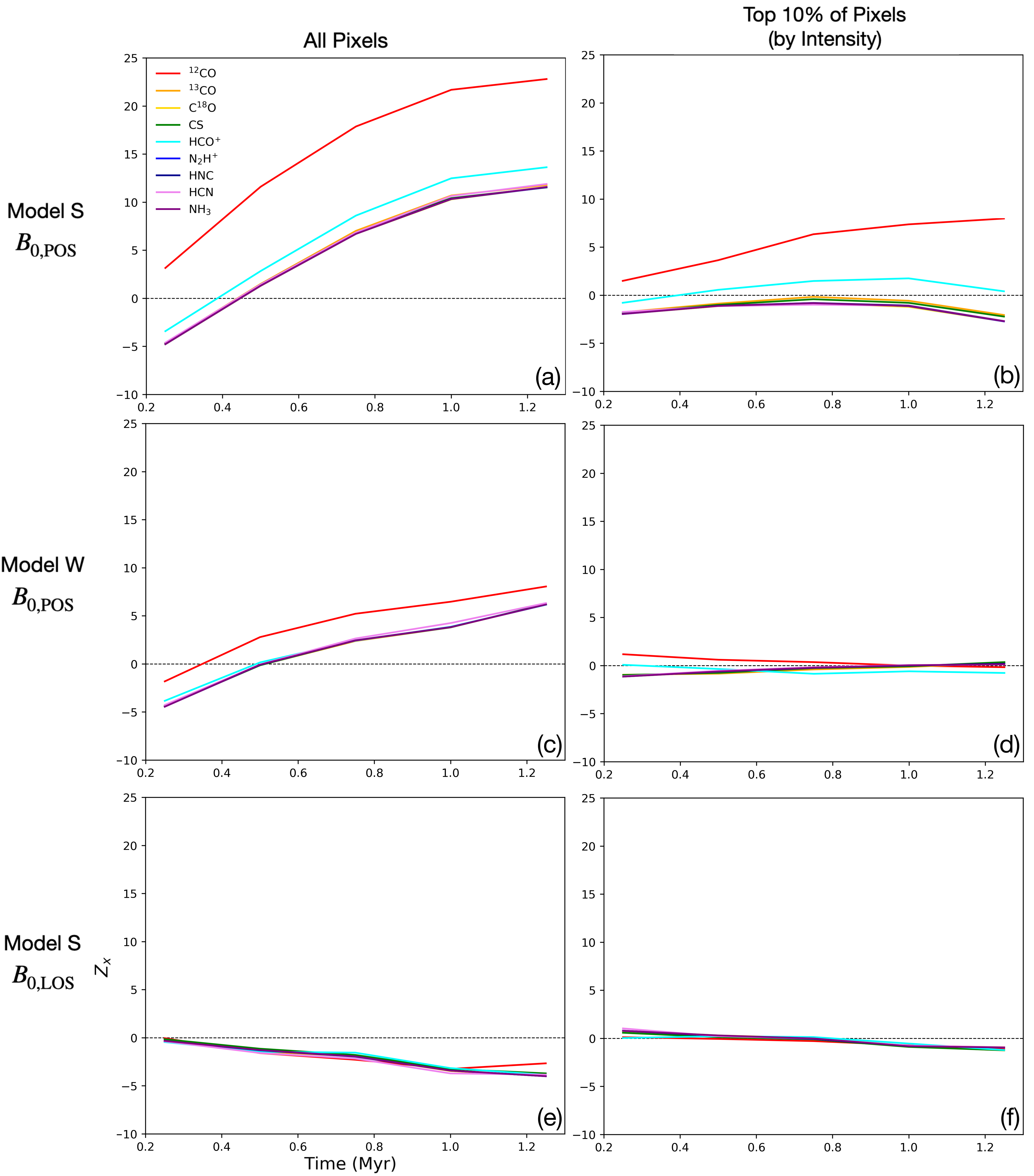}
\caption{The Projected Rayleigh Statistic ($Z_X$) as a function of time for each of our tracers, for a variety of scenarios within our main grid of models. The left column shows the results using the whole observer space, and the right column incorporates a cut that includes only the top 10\% highest intensity pixels in the computation. \textit{\textbf{Panels (a) and (b):}} Fiducial model results (stronger magnetic field simulation, viewed with $B_0$ in the plane-of-the-sky). \textit{\textbf{Panels (c) and (d):}} Same as fiducial model, but with the weaker field simulation (Model W). \textit{\textbf{Panels (e) and (f):}} Same as fiducial model, but viewed with $B_0$ along the line-of-sight.} \label{fig:prs}
\end{figure*}

Here we compare the results obtained for Model S (the fiducial model) and Model W, which has a factor of four weaker magnetic field strength.
Panels (c) and (d) of Figure \ref{fig:prs} show the PRS vs. time for each of these simulations.
There is a substantial difference in alignment preference between the two models, especially when only the highest intensity cells are considered.
Considering all pixels of the synthetic maps (left panels in Figure \ref{fig:prs}), both models show a tendency toward higher $Z_x$ at later times, as the simulation becomes more gravitationally dominated. 
This effect is mainly leveraged on the lower intensity regions, however, as discussed in the previous section.
In the top 10\% of cells case, the difference between $^{12}$CO and the high density tracers is absent in Model W, except at the very earliest times when the magnetic field is still roughly uniform (as prescribed by the initial conditions).
As the simulation progresses, the weaker magnetic field of Model W is tangled by turbulent gas flows and its orientation becomes essentially random.
As a result, none of the HROs have any correlative preference.
This result shows the utility of this method in diagnosing magnetic field strength, as alignment preference between polarization vectors and molecular structure (and moreover, parallel vs. perpendicular preference in low- vs. high-density tracers) only appears when the magnetic field is sufficiently strong to have dynamical importance.

It is also notable that the parallel alignment observed in the "all cells" cut (which contains mostly low-intensity cells) persists for both the strong and weak field simulations.
In both cases, the polarization vectors in the low-intensity regions are dominated by contributions from low-density gas, where density structures tend to lie parallel to the plane-of-the-sky magnetic field orientation.
For $^{12}$CO, this is the case even in pixels that include high-density gas (see the pink curves in Figure \ref{fig:hros}).
These results suggest that, generally, the bulk orientation of the magnetic field in diffuse regions of the cloud may be inferred just from the intensity gradients.
Therefore, methods that rely solely on this information, such as the intensity gradient technique \citep{hu2019b}, should have significant power in mapping out the magnetic field in the outer layers of a cloud, even when the magnetic field is not particularly strong.
Of the tracers we tested, $^{12}$CO is the best option for this task since it explicitly probes the low-density gas due to optical depth effects (see Section \ref{ssec:depthDisc}).

\subsection{Inclination Effects}\label{ssec:inclination}

Since dust polarimetry only probes the plane-of-the-sky component of the magnetic field, viewing orientation has an impact on observed HRO outcomes.
Furthermore, if the magnetic field is strong enough to regulate gas flows then the observed gas structure will also be affected.

First, we can compare the two inclination extremes - the $B_{0,\rm POS}$ case (i.e., the fiducial model, with $i=90^{\circ}$) versus the $B_{0,\rm LOS}$ case ($i=0^{\circ}$).
The PRS as a function of simulation time for these cases are plotted in panels (a)-(b) and (e)-(f) of Figure \ref{fig:prs}, respectively.
Whereas the magnetic field in the plane-of-the-sky orientation showed significant secular evolution of the PRS and different alignment preference for the low- and high-density tracers (in the the "all cells" version and the top 10\% cut), these effects are entirely suppressed when the bulk field is along the observer's line-of-sight.

\begin{figure}
\centering
\includegraphics[width=\columnwidth]{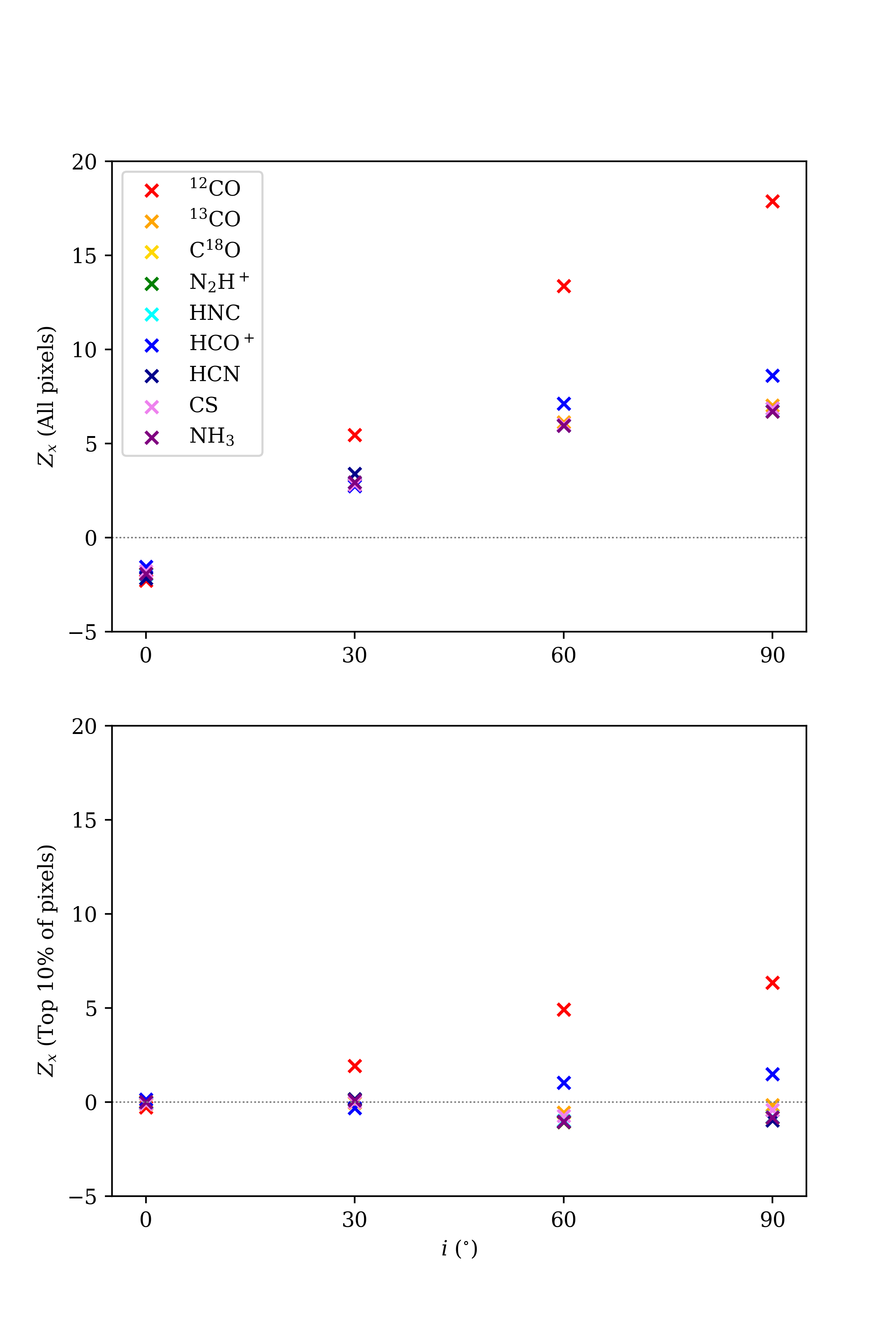}
\caption{The Projected Rayleigh Statistic results for each of our tracers for a variety of magnetic field inclinations, using the fiducial $t=0.75$ Myr time step of Model S. An inclination of $i=90^{\circ}$ corresponds to the $B_{0,\rm POS}$ view, and $i=0^{\circ}$ corresponds to the $B_{0,\rm LOS}$ view. The top and bottom panels show the results for all pixels in the observer space and only the top 10\% of pixels, respectively. } \label{fig:barinter}
\end{figure}

For the fiducial timestep, we also computed $Z_x$ for each of our tracers at a few intermediate inclinations ($60^{\circ}$ and $30^{\circ}$) using Model S.
The results of this exploration are presented in Figure \ref{fig:barinter}.
The outcome is largely consistent with expectation for a mostly rigid magnetic field.
For the calculation that considers all cells of the simulation (i.e, is dominated by the low-density regions, where we expect more parallel alignment), the PRS decreases roughly linearly as the plane-of-the-sky component of $\boldsymbol{B}$ decreases. 
The PRS stays above zero for all tracers even down to $i=30^{\circ}$, and the value for $^{12}$CO remains notably larger than the values for the high density tracers.
There is a different outcome when we only consider the top 10\% of cells. 
$^{12}$CO follows roughly the same trend, but the other tracers (with the exception of HCO$^{+}$) show moderately negative $Z_x$ for $i = 90^{\circ}$ and $i = 60^{\circ}$.
This is because with this 10\% cut we only include the highest column density cells, which have a preference for perpendicular alignment.
However, as the view becomes more inclined (and the plane-of-the-sky component of the magnetic field is thereby reduced), this preference is eliminated and $Z_x$ moves toward zero.
Interestingly, even at $i = 30^{\circ}$ there is essentially no perpendicular alignment in any of the tracers, other than $^{12}$CO.

\section{Discussion}\label{sec:discussion}

Our discussion is split into four parts.
In Section \ref{ssec:depthDisc} we use optical depth information to connect our results to the 3D picture of our simulations.
We then comment on the possible physical implications of a given observation as implied by our simulations. 
In Sections \ref{ssec:caseStudy} and \ref{ssec:lvgComp}, we examine the impact of varying the abundance (for a given molecular tracer) and radiative transfer assumption (LTE vs. LVG), respectively.
Finally, in Section \ref{ssec:blastComp} we apply beam convolution to selected synthetic observations and investigate the corresponding HRO results, to provide a more direct comparison with Vela C data \citep{fissel2019}.  

\subsection{Optical Depth Connection}\label{ssec:depthDisc}

\begin{figure}
\centering
\includegraphics[width=\columnwidth]{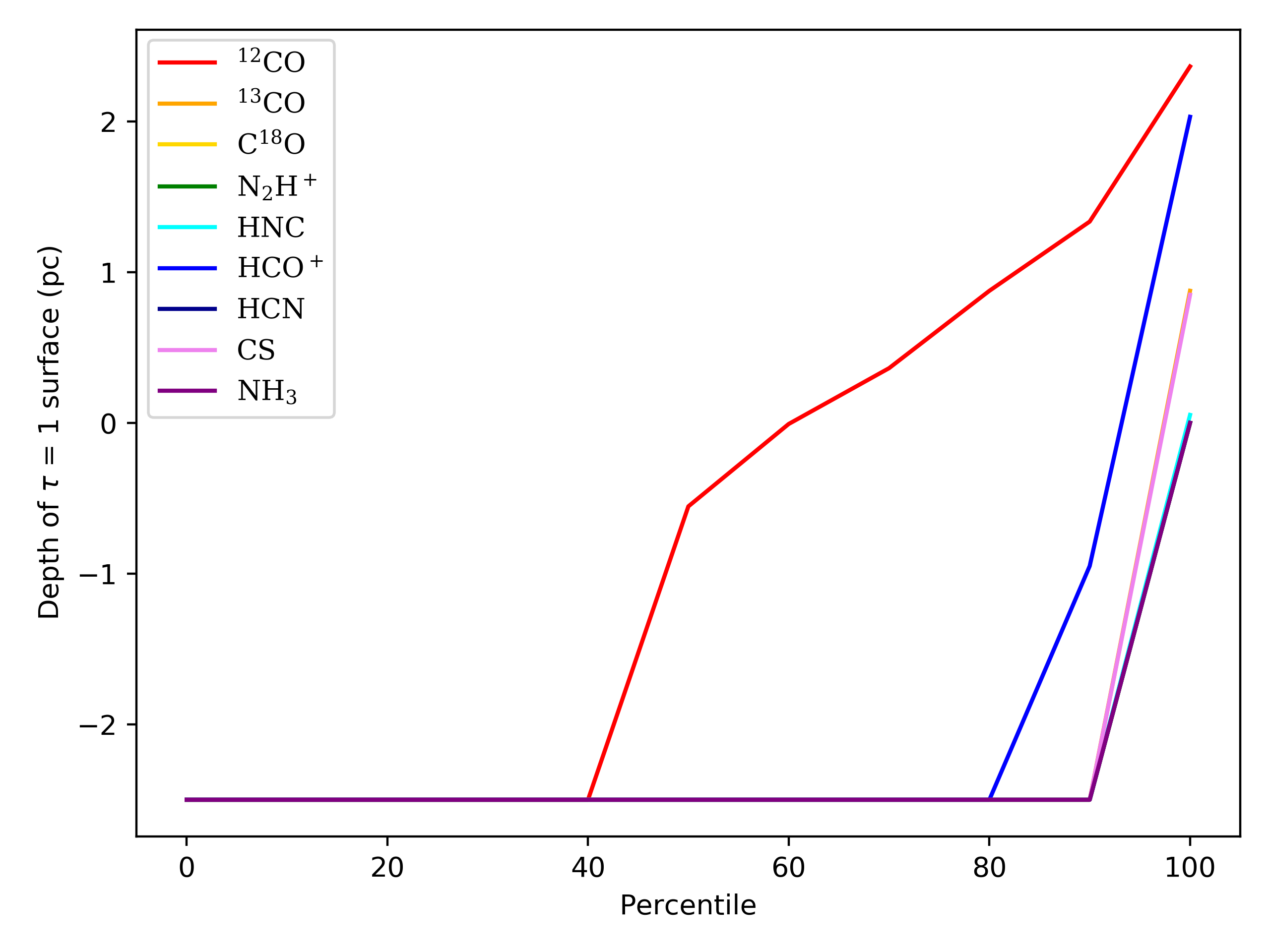}
\caption{The depth of the $\tau = 1$ surface for each of our tracers, as calculated from the fiducial frame in Model S. Whereas for $^{12}$CO $\sim$40\% of pixels are fully optically thin (corresponding to the very lowest intensity pixels, in the outskirts of the cloud), all the other tracers are optically thin in $\gtrsim$90\% of pixels (with the exception of HCO$^{+}$, at $\sim$80\%). For $^{12}$CO the $\tau = 1$ surface is encountered in front of the midplane in about 40\% of pixels. Note: in this figure the line for C$^{13}$O is overlapped by the CS line, and the lines for N$_2$H$^+$, HNC, C$^{18}$O, and HCN are overlapped by the NH$_3$ line.} \label{fig:tau1surfall}
\end{figure}

To facilitate this analysis, we used RADMC3D to calculate the physical depth of the $\tau = 1$ surface for each of our observed molecular tracers (at line center).
Results taken from the fiducal observing frame from Model S are shown in Figure \ref{fig:tau1surfall}.
For many of the high density tracers (NH$_3$, HCN, HNC, N$_2$H$^+$, C$^{18}$O), almost the entire cloud along the line-of-sight is optically thin; the $\tau = 1$ surface is not encountered between the observer and the far side of the cloud for 90\% of sightlines, and for all the remaining 10\% of sightlines the $\tau = 1$ is reached only beyond the midplane.
For CS, $^{13}$CO, and HCO$^{+}$, about 5\% of sightlines (10\% for HCO$^{+}$) encounter $\tau = 1$ just in front of the midplane.
These regions are well correlated with the highest column density sight-lines.

\begin{figure}
\centering
\includegraphics[width=\columnwidth]{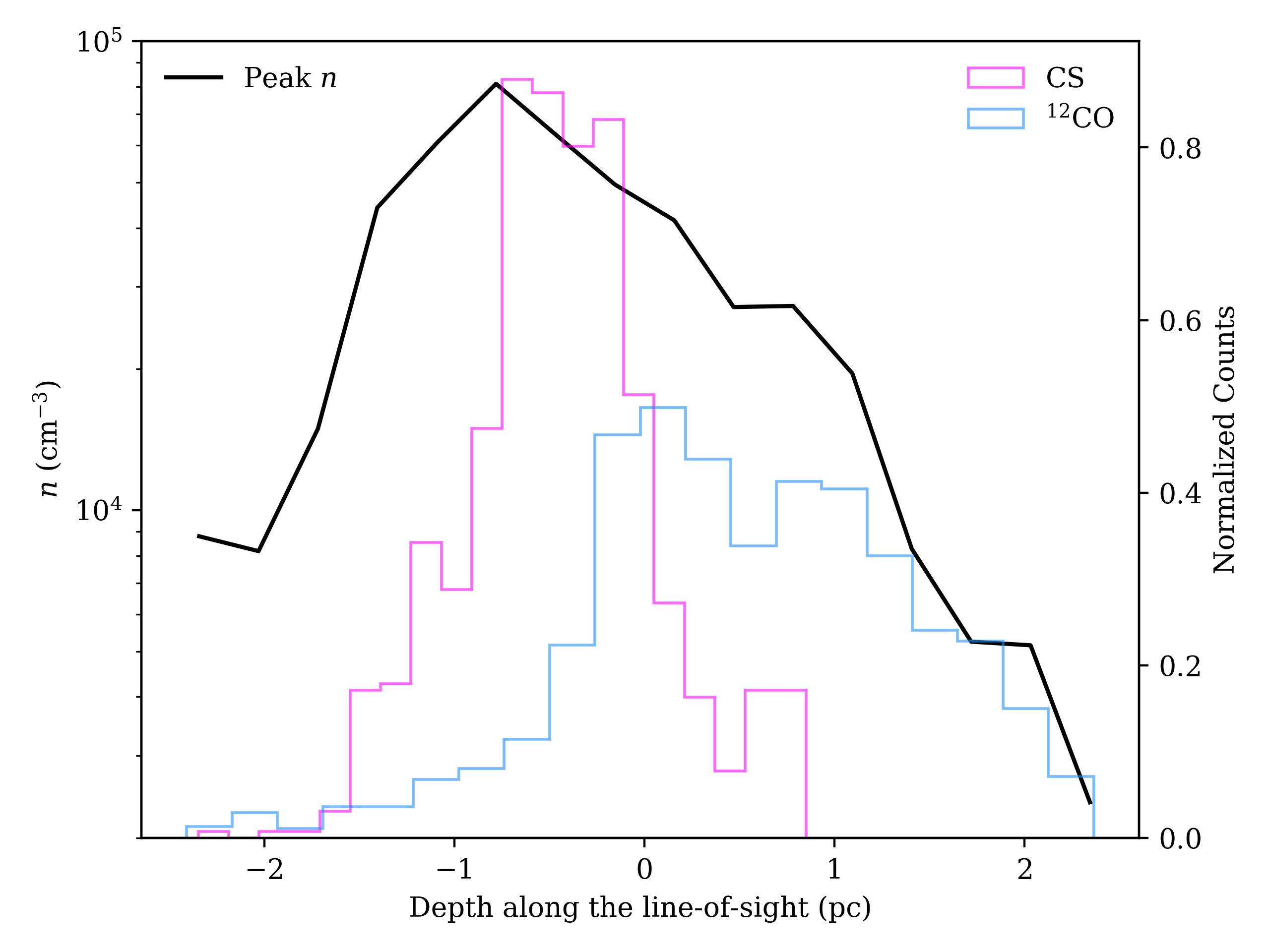}
\caption{Comparison between the maximum volume density reached along the observer's line-of-sight and the histograms of the $\tau = 1$ surface depths for $^{12}$CO and CS. The observer is placed at positive infinity on the axis. The CS distribution is well-aligned with the location of the highest volume density gas, consistent with the expectation that CS operates as a high-density tracer. Meanwhile, the $^{12}$CO is offset toward the front of the cloud where $n$ is lower, confirming that it largely probes the low density gas.} \label{fig:tau1cocs}
\end{figure}

$^{12}$CO is the most optically thick tracer by a wide margin.
For about 40\% of the observer space the $\tau = 1$ surface is encountered in $^{12}$CO before the midplane.
These high opacity sightlines overlap with some of the highest column density regions in the simulation (see Figure \ref{fig:tau1cocs}), thus obstructing the observer's view of high volume density regions.
This provides physical evidence for why $^{12}$CO produces a distinct alignment trend from the others.
That is, the data from $^{12}$CO are clearly probing a different section of the cloud, only receiving emission from the lower density foreground along the optically thick sight lines.
The "high density" tracers also probe this low-density material, but since they are optically thin up to (and in some cases, through) the midplane of the cloud, the final intensity ultimately becomes dominated by the high density gas.

To demonstrate this effect, plotted in Figure \ref{fig:tau1cocs} are histograms of the location along the line-of-sight of the $\tau = 1$ surface for tracers that are representative of these two regimes, $^{12}$CO and CS.
On a dual axis, we compare this to the maximum number density reached in a slice of the cloud (parallel to the plane-of-the-sky) at a given depth into the cloud along the line of sight. 
From this visualization, we can see that the peak of the distribution for CS is nearly aligned with the maximum volume density in the simulation box. 
The observed CS emission therefore serves as a good probe of the high density gas in the simulation, and it is able to trace the high density filaments visible in the column density maps (see e.g., Fig \ref{fig:fidMoment0}).
As a result, the perpendicular alignment between 3D gas structure and magnetic fields at the highest volume densities (see e.g., Figure \ref{fig:cheyuplots} panel (a)) is preserved in the high intensity cut for CS (negative $Z_x$).
Meanwhile, the $^{12}$CO $\tau = 1$ surface distribution is offset from the peak volume density along the line-of-sight toward the closer side of the cloud to the observer.
Therefore, the observed $^{12}$CO emission only probes (on average) lower density gas.
This material shows a preference for parallel alignment, which is reflected in the $^{12}$CO HRO results.

Notably, the results found here are consistent with those obtained in a similar experiment by \citet{hu2019a}.
They studied the effectiveness of 
velocity channel gradients \citep[VChGs; see, e.g.,][]{lazarian2018} in probing the magnetic field orientation using molecular tracers $^{12}$CO, $^{13}$CO, C$^{18}$O, CS, HNC, HCO$^{+}$, and HCN.
They found that $^{12}$CO, and to lesser extent $^{13}$CO, is much more effective at reconstructing the magnetic field in low-density regions than in high-density regions.
In contrast, lower optical depth tracers (such as  C$^{18}$O, CS, and HNC) were effective over a larger range of densities.
This result lead to the suggestion that $^{12}$CO can serve to study the field structure in the outer layers of the cloud, and the lower optical depth tracers to trace high-density structure.
Our work here supports this idea, with Figure \ref{fig:tau1surf} especially showing that it is indeed the case that line radiative transfer observations of $^{12}$CO trace different (outer layer) gas as compared to the tracers that probe more deeply into the high-density material in the cloud midplane.

\subsection{CS and $^{12}$CO Abundance Case Studies}\label{ssec:caseStudy}

The previous section discussed different gas-magnetic field alignment preferences in synthetic molecular line observations, and we attributed such difference to the different density ranges traced by these molecules. The synthetic observations, however, also depend on species abundances, 
which in our main grid of models we prescribed using observational constraints (see Table \ref{table:fidAbundances}).
We therefore performed a case study of varying species abundance to explore how it may affect synthetic line emissions.

\begin{figure}
\centering
\includegraphics[width=\columnwidth]{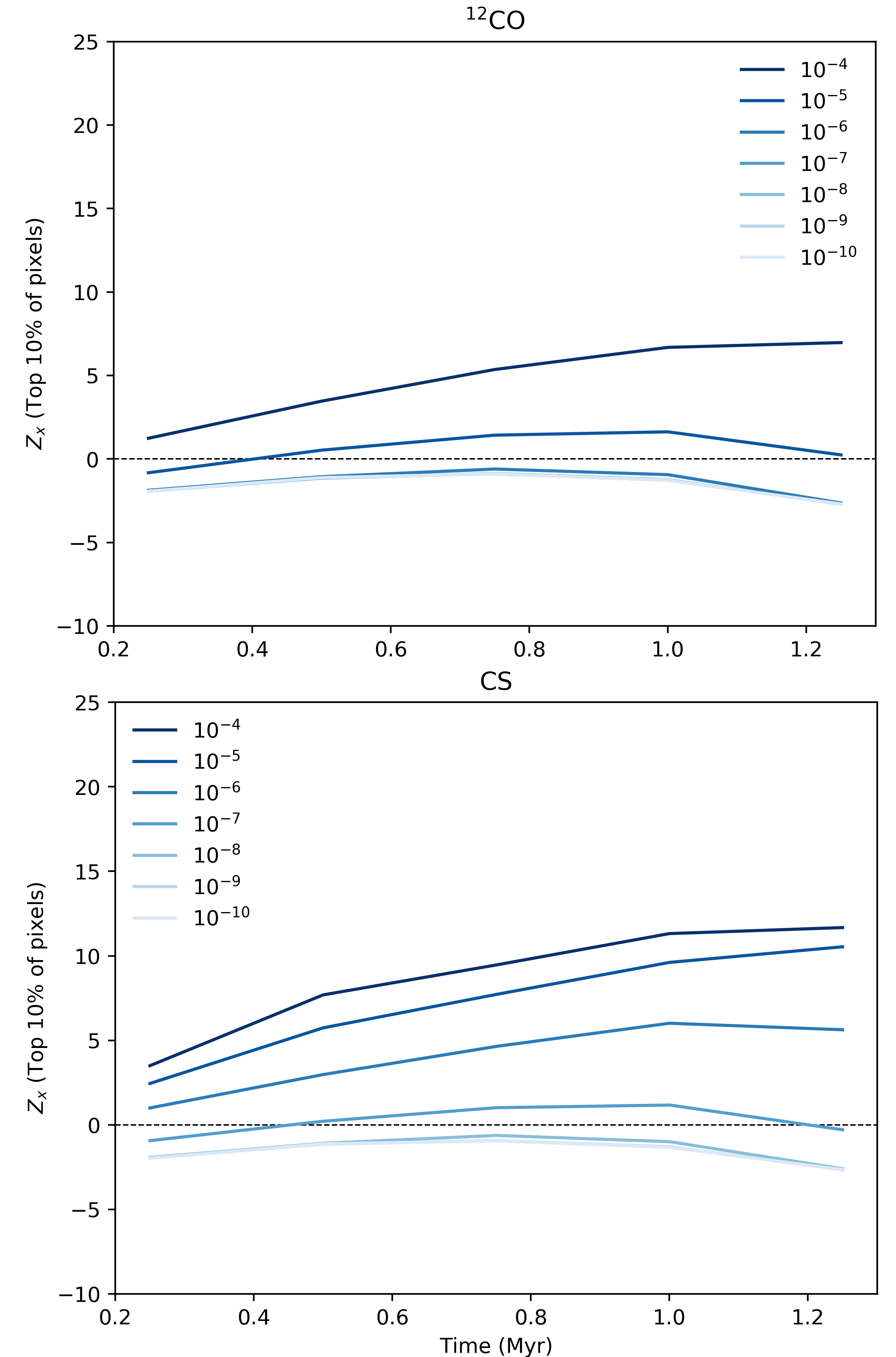}
\caption{The Projected Rayleigh Statistic calculated using the top 10\% highest intensity cells for $^{12}$CO (top) and CS (bottom) at abundances ranging from 10$^{-10}$ to 10$^{-4}$. At very low abundance both tracers exhibit perpendicular alignment. As the value is increased, they both move to positive $Z_x$, due to the corresponding increase in optical depth. In cloud environments CS and $^{12}$CO are observed to have abundances of $\sim$10$^{-8}$ and $\sim$10$^{-4}$, respectively, so they are typically associated as high- and low-density tracers, respectively.} \label{fig:abunAdjust}
\end{figure}

Shown in Figure \ref{fig:abunAdjust} are the PRS results for synthetic observations of $^{12}$CO and CS with abundances ranging from $10^{-10}$ to $10^{-4}$.
Interestingly, the PRS values from a single molecular line with various abundances highly resemble those from various gas tracers (see panel (b) of Figure \ref{fig:prs}). 
Note that the $^{12}$CO and CS show nearly identical results at very low abundance.
This represents the optically thin observing scenario, and as such the particular radiative transfer properties of the tracer are irrelevant - the observer sees all of the emission, essentially reproducing the column density map simply calculated by summing up the volume density along the line-of-sight (Equation \ref{eq:cd}).
However, when the abundance value is turned up, optical depth begins to come into play and high column density material becomes obscured.
The transition abundance is different ($\sim$10$^{-5}$ for $^{12}$CO, and $\sim$10$^{-7}$ for CS), but in both cases the $Z_x$ value systemically increases after that threshold is reached.

This case study demonstrates that optical depth is the critical factor controlling the observed gas-magnetic field alignment preference. 
In other words, operative quantity that $Z_x$ depends on is the density of cloud material being probed, which ultimately is principally a function of the depth into the cloud probed by the observation.

\subsection{LTE vs. LVG}\label{ssec:lvgComp}
\begin{figure*}
\centering
\includegraphics[width=\textwidth]{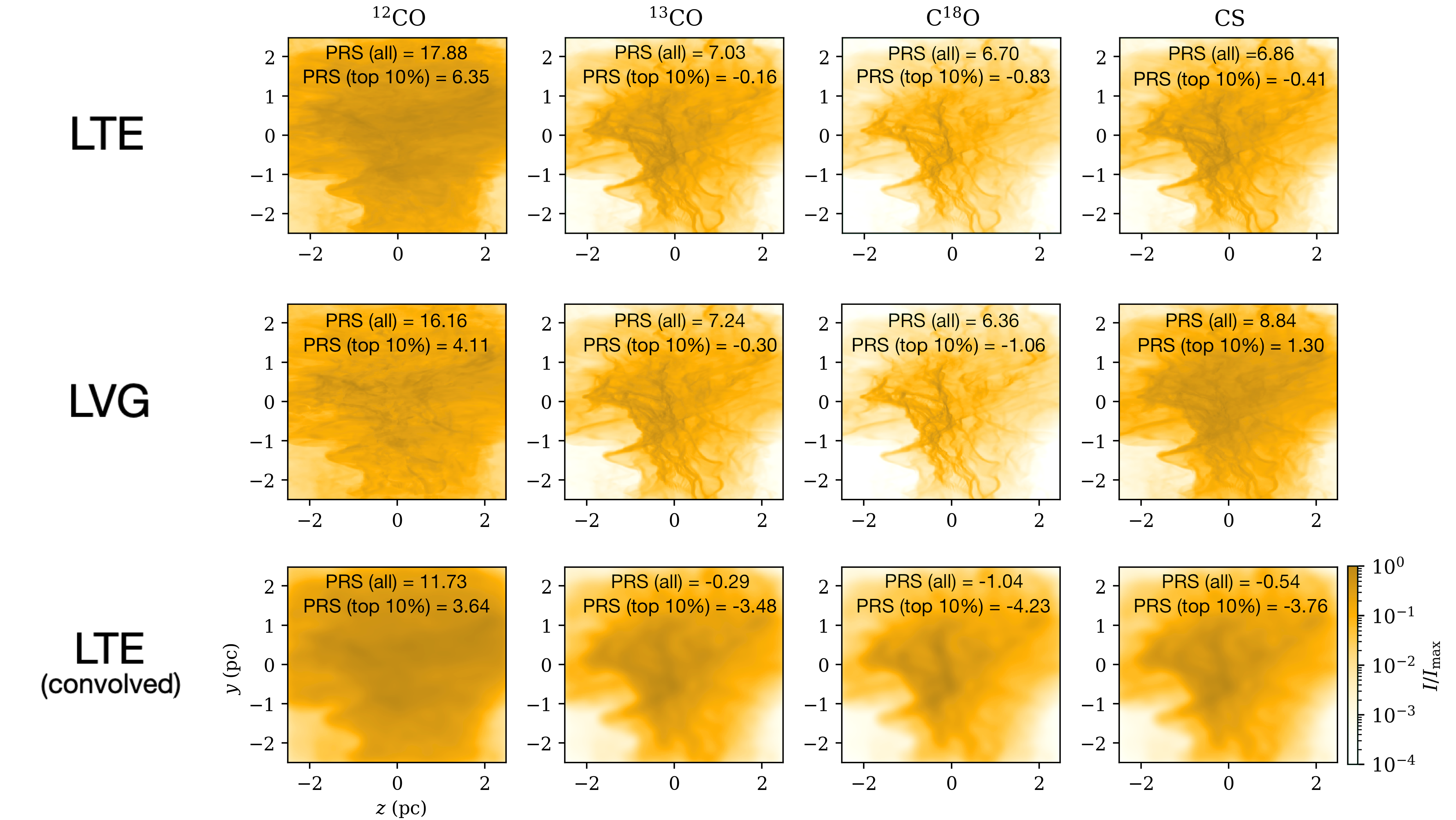}
\caption{The effects of changing the radiative transfer mode (middle row: LVG) and beam size (bottom row: 0.2 pc FWHM Gaussian convolution) on (normalized) Moment 0 maps for a sub-set of the molecules we simulated. Note that the switch to LVG is not uniform across the tracers; it has obvious effects on the $^{12}$CO and CS emission, but there is virtually no change for ${13}$CO and C${18}$O.} \label{fig:discMaps}
\end{figure*}

Our main results were computed using the LTE assumption, with $T = 10$ K set to match the isothermal conditions prescribed in our MHD simulations.
In the top two panels of Figure \ref{fig:discMaps}, we compare these results to those obtained using another commonly-adopted radiative transfer mode, the Large Velocity Gradient (LVG) approximation, for a sub-selection of our tracers.
The $^{12}$CO emission becomes somewhat more patchy, compared to the fairly uniform LTE case.
This is expected, since at each point in the simulation space the LVG method uses a calculation of gas velocities near a given location to determine photon escape probabilities, which produces different results at different locations.
Interestingly the patchiness in the $^{12}$CO introduces some perpendicularly aligned structure, slightly lowering the PRS.
Even so, the overall parallel-alignment character of the $^{12}$CO remains.

\begin{figure*}
\centering
\includegraphics[width=\textwidth]{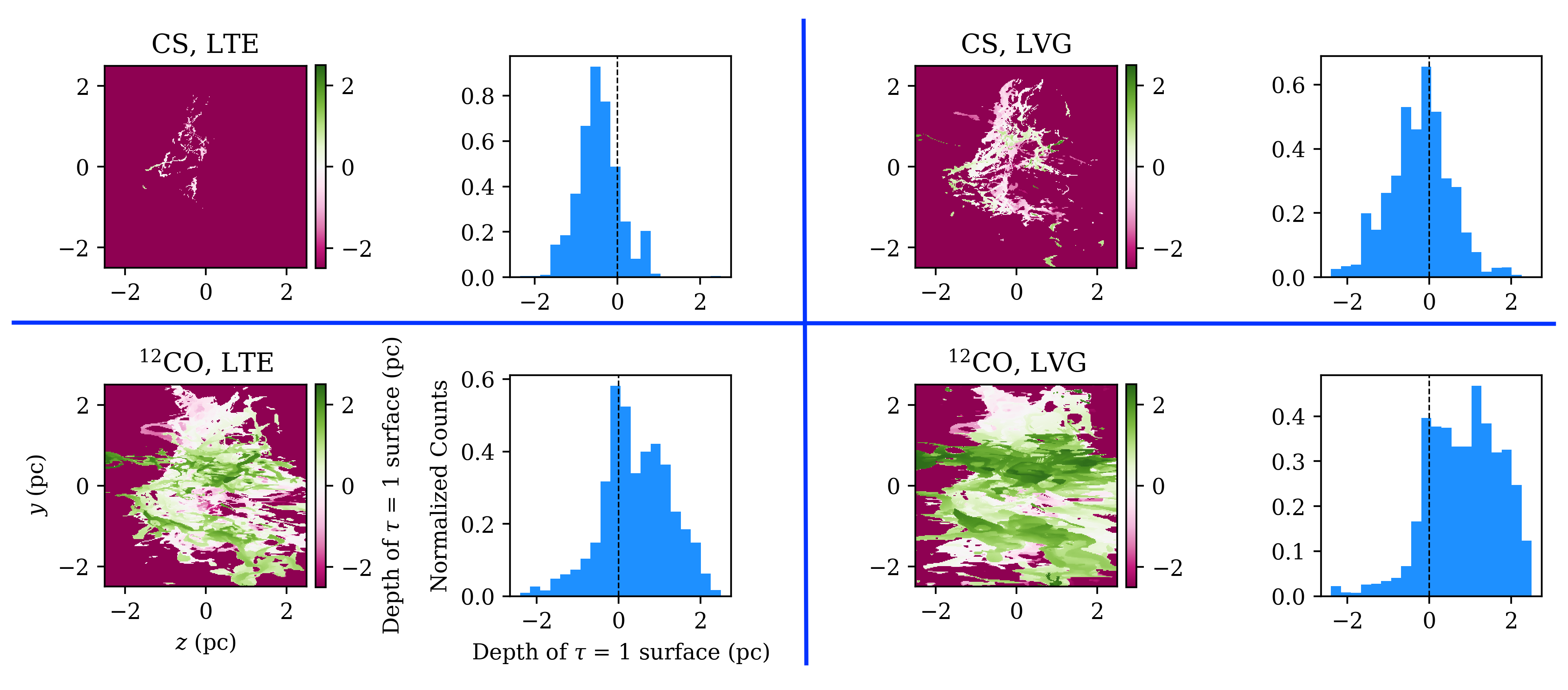}
\caption{Comparison of the depth of the $\tau = 1$ surface in LTE vs. LVG, for $^{12}$CO and CS. In both cases, switching the radiation transfer mode to LVG brings the $\tau = 1$ surface closer to the observer on average. This translates to a higher $Z_x$ in the CS, as it transitions from optically thin to optically thick in some sub-regions.} \label{fig:tau1surf}
\end{figure*}

The effect of changing from LTE to LVG differs among the high-density tracers. 
The CO isotopologues ($^{13}$CO and C$^{18}$O) both show very little change, both visually and in terms of the PRS values.
However, the CS produces a significantly different result, with the LVG map becoming much more opaque and revealing less of the high density filamentary material.
In turn, this leads to an increase in the PRS, with the densest sightlines (top 10\%) flipping from overall perpendicular alignment to parallel alignment.
The top row of Figure \ref{fig:tau1surf} shows the impact on the CS optical depth.
The LVG calculation produces a $\tau = 1$ surface that is on-average closer to the front of the cloud, with respect to the observer. 
This agrees well with the analysis performed above in Section \ref{ssec:depthDisc}.
Of note, the $^{12}$CO $\tau = 1$ surface also moves closer to the observer when LVG is employed.
This does not translate to an increase is $Z_x$, however, because even in LTE the $^{12}$CO was already quite opaque.

We found that in LTE, the occupation of the upper and lower levels of the of the $J = 1-0$ transition for CS were 25\% and 11\%, respectively.
In LVG this increased to 44\% and 28\%, which is comparable to the LTE values for the $^{12}$CO $J = 1-0$ transition (43\% and 25\%).
This higher occupation, particularly in the upper level, leads to more emission and, in the case of CS, an increase in the optical depth of the transition.

Here, we have briefly highlighted that the physical method for simulating the radiative transfer can have some impact on the results for a given molecular tracer, with the operative effect being that the redistribution of level populations can produce emission with higher optical depth if the occupation of the upper level of the transition is increased (or lower optical depth if the occupation of the upper level is decreased). 
The overall physical interpretation of the PRS and its relationship to opacity (i.e., that optical depth is a key driver of the $Z_x$ value, as it dictates whether the high density filaments are visible) remains unchanged.
Future work may include a more detailed exploration of the specific impact of radiative transfer mode on the HRO results for a wider variety of physical situations and tracers.

\subsection{Beam Convolved Results \& Comparison with BLASTPol Vela C Observations}\label{ssec:blastComp}
In \citet{fissel2019}, BLASTPol polarization results and Mopra molecular line observations were used to calculate the PRS for the Vela C molecular cloud.
Though our MHD simulation setup was not specifically tailored to Vela C (rather, it is meant to be a generic collapsing cloud), it can still be useful as a point of comparison for the results in \citet{fissel2019}.
To generally replicate observational conditions, we convolve our (LTE) line radiative transfer intensity maps and polarization maps with a variety of Gaussian beam sizes ranging from 0.2 to  0.6 pc (FWHM), then recompute the HRO and PRS. 

\begin{figure*}
\centering
\includegraphics[width=\textwidth]{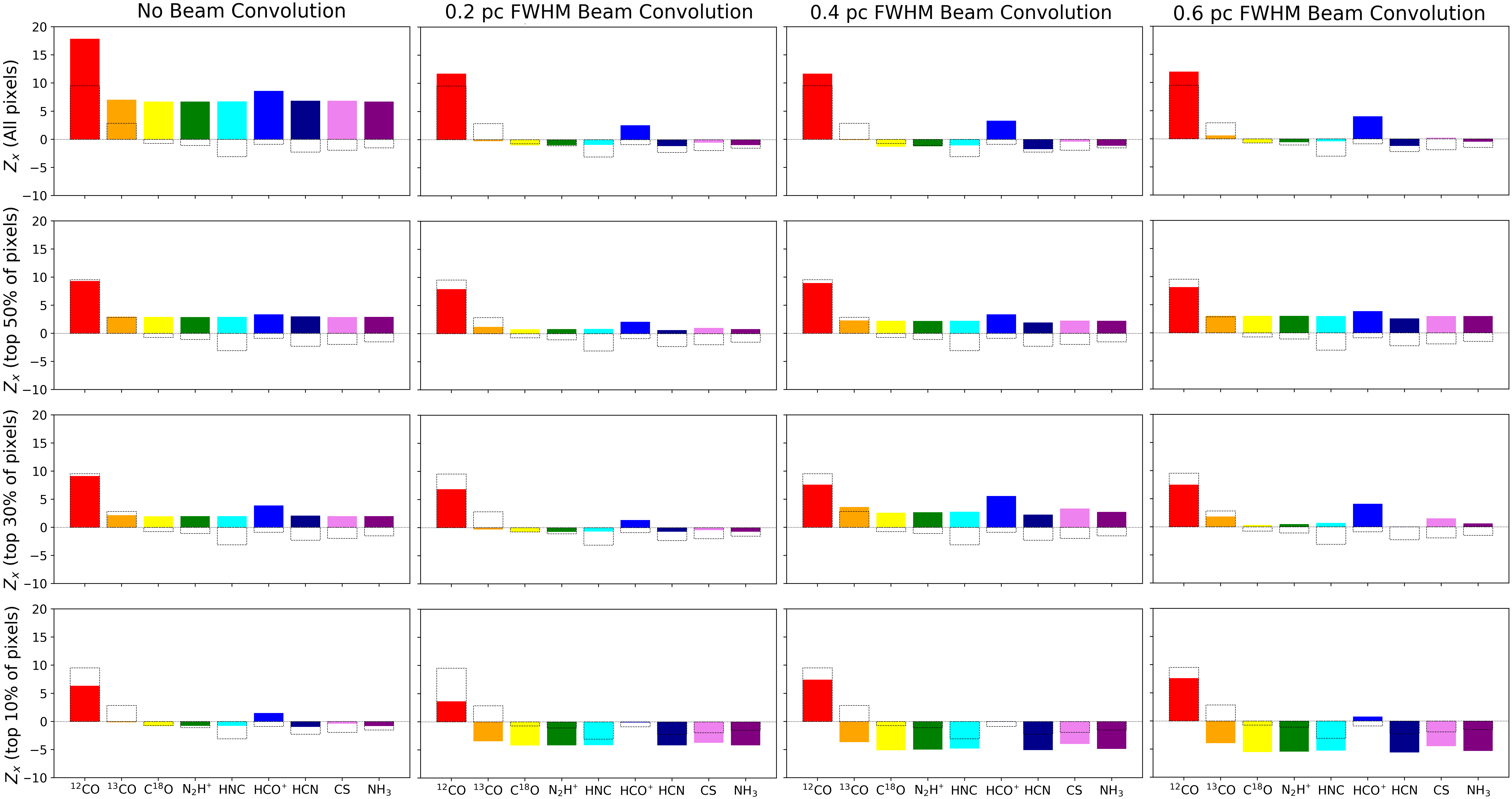}
\caption{PRS results from the fiducial frame for Model S, for a matrix of Gaussian beam sizes and sensitivity cuts. For comparison, the PRS values for the Vela C molecular cloud computed by \citet{fissel2019} are included in each plot (as dashed bars). Generally, the Vela C results for $^{12}$CO and $^{13}$CO are best reproduced by our lower sensitivity models (top 50\% or all pixels), and the results for the remaining tracers (all of which have negative PRS values in Vela C) are best reproduced by our lower sensitivity models (top 30\% or top 10\% of pixels). Beam convolution tends to decrease the value of the $Z_x$, especially in the highest intensity pixels.} \label{fig:conv2}
\end{figure*}

These data are presented in Figure \ref{fig:conv2}, alongside a comparison to the PRS results obtained from Vela C. (adapted from \citet{fissel2019}).
In addition to the 10\%-intensity cut used for the main set of models, we also present results with softer cuts of 30\%, and 50\% of sightlines (see Table \ref{table:sensCuts} for cutoff values).
This is designed to resemble the sensitivity limitations in real observations.
These results show that beam convolution does have some impact on the outcome of the HRO calculations.
Particularly, it tends to reduce $|Z_x|$ across all tracers, especially for the hardest (10\%) and softest (50\%) cuts.
In the 50\% case, it is possible that this effect is partially enhanced by the fact that the beam convolution tends to emphasize the edge of the cloud (this can be seen visually in the bottom row of Figure \ref{fig:discMaps}).  

\begin{table}
\caption{
The values corresponding to the 50th, 70th, and 90th percentile intensity cuts for each molecular tracer simulated in this work, as taken from the fiducial frame in Model S. \label{table:sensCuts}
}
\begin{tabular}{cccc}
\hline
Species & 50th (K km/s) & 70th (K km/s) & 90th (K km/s)\\
\hline
$^{12}$CO & 12.84 & 18.91 & 28.64 \\
$^{13}$CO & 2.29 & 5.22 & 13.17 \\
C$^{18}$O & 0.70 & 1.70 & 4.66 \\
N$_{2}$H$^+$ & 0.26 & 0.63 & 1.70 \\
HNC & 0.41 & 0.96 & 2.59 \\
HCO$^{+}$ & 6.18 & 12.54 & 25.09 \\
HCN & 0.33 & 0.78 & 2.11 \\
CS & 1.89 & 4.40 & 11.40 \\
NH$_3$ & 0.41 & 1.04 & 2.81 \\
\hline
\end{tabular}
\end{table}

Our results for the multiple intensity cuts provide a potential explanation for the wide range of $Z_x$ computed for the different tracers in Vela C. 
Higher abundance molecular like $^{12}$CO and $^{13}$CO tend to be observable across the whole cloud, whereas the lower abundance ("high-density") molecules require greater sensitivity, and can only reach the sensitivity threshold in the highest column density regions of the observer space.
Accordingly, the $Z_x$ computed for softer cut (50\%) is in reasonably good agreement with the Vela C results for $^{12}$CO and $^{13}$CO, and the high-density tracers agree better when the calculation is restricted to only the top 10\% of pixels.
It is worth emphasizing again, however, that our simulation is not tailored to mimic Vela C, and some of the discrepancy between the observational and synthetic results may also be due intrinsic physical differences between the physical set-up of our model and the underlying physical conditions in Vela C. 
More broadly, this experiment demonstrates that the sensitivity of an observation can have an important effect on the observed HRO results.
The very brightest pixels correspond to the highest column density, where structures preferentially align perpendicular to the magnetic field.
Therefore, lower sensitivity observations will tend to bias the results toward negative $Z_x$, because the lower column density regions are excluded.

One additional important caveat for this analysis is that all of the radiative transfer simulations performed in this work assumed uniform relative abundance.
A variable prescription, particularly one that assumes higher relative abundance at higher volume density, would likely produce results that further emphasize the perpendicular character of the high-density tracers.
Likewise, in the very lowest density regions of our simulation box (including the ambient gas outside the main cloud) there is no taper in the abundance.
This is probably unrealistic, as molecular gas does not form as readily in the lower density ISM.
Therefore, our results that only use 50\% of the pixels may be generally more useful for observational comparison than the results using all the pixels.

\section{Conclusions} \label{sec:conc}
In this work, we performed synthetic dust polarization and molecular line observations of 3-dimensional MHD data to investigate the relationship between the magnetic field and gas density structure in a collapsing molecular cloud using the HRO technique.
We observed our cloud at multiple stages of evolution over $\sim$1 Myr, and from a variety of viewing geometries. 
Additionally, we tested how the results were affected by changing the global magnetic field strength, the abundance of our selected species, and the chosen radiative transfer mode (LTE vs. LVG). 
To link our theoretical exploration to the current observational state of the field, we also compared selected results with the analysis of Vela C performed by \citet{fissel2019}.
Our main conclusions are as follows:

\begin{enumerate}
    \item Analysis of our magnetized turbulent cloud simulations shows that the relative orientation between the 3D magnetic field and density structures depends on the cloud-scale magnetic field strength (see Figure \ref{fig:cheyuplots}). The super-Alfv\'enic simulation (Model W) showed significantly more magnetic field tangling, and as a result had relatively little alignment preference compared to the trans-Alfv\'enic simulation (Model S), especially at high densities. In Model S the mean magnetic field remained relatively rigid, leading to the formation of orthogonal elongated filaments. In the lower density regions, thin streams of gas formed along the the field lines (Figure \ref{fig:streams}). These structures are visible in slice plots of our simulations, and are similar to the striations seen in observations. In Model S, the transition from parallel alignment to perpendicular alignment occurs at $n \gtrsim 4 \times 10^{3}$ cm$^{-3}$ for the $t = 0.75$ Myr timestep. Notably, this is in close agreement with the transition value of $\sim$1000 cm$^{-3}$ estimated for Vela C \citep{fissel2019}.
    \item The intrinsic relative gas-magnetic field orientations observed in 3D are also visible in the HRO analysis using our synthetic molecular line observations (Figure \ref{fig:hros}). Excluding the very earliest times, the PRS results with no intensity cut ("all pixels") generally show (for both Model S and Model W) parallel alignment when the mean magnetic field is in the plane-of-the-sky. This is because the overall observer-space is dominated by low density sightlines. The situation changes when we consider only the highest intensity (top 10\%) pixels. For Model W, each of the tracers show a PRS value close to zero in this cut. Meanwhile, for Model S there is some stratification; most show a $Z_x \lesssim 0$ across all times, however $^{12}$CO has a positive value.  
    \item In the $B_{\rm POS}$ view (i.e., the viewing orientation with the mean magnetic field in the plane-of-the-sky), the difference in HRO result between $^{12}$CO and the high density tracers is closely linked to optical depth effects. The $\tau = 1$ surface for $^{12}$CO is on average closer to the front of the cloud (from the observer's point of view) than the $\tau = 1$ surface for CS (Figure \ref{fig:tau1surf}). This results in the $^{12}$CO emission mainly probing the low-density gas where alignment preference is dominated by the magnetic field-aligned striations, instead of the denser filaments which align orthogontal to the mean magnetic field. This is the case even in the highest intensity pixels. Though this discussion is primarily framed as a distinction between molecular lines, it is also the case that if we drastically vary the abundance for a given tracer, both positive $Z_x$ and negative $Z_x$ results can be produced (Figure \ref{fig:abunAdjust}). This demonstrates that for a given observation it is the optical depth itself that is the essential quantity in determining which regions of the cloud are probed.  
    \item When the mean magnetic field is inclined away from the plane-of-the-sky (we define the $B_{\rm POS}$ view to have $i=90^{\circ}$ in this work), the absolute value of $Z_x$ tends to decrease (Figure \ref{fig:barinter}). At intermediate inclinations of $i=60^{\circ}$ and $i=30^{\circ}$, there is still moderate differentiation between the PRS for $^{12}$CO and the high-density tracers, but this vanishes when $i \rightarrow 0^{\circ}$. This produces some degeneracy between the results for a weak magnetic field (Model W) and a strong magnetic field oriented along (or nearly along) the line-of-sight.
    \item Our main set of results used synthetic line radiative transfer observations that were generated assuming LTE at $T = 10$ K. For a subset of our tracers ($^{12}$CO, $^{13}$CO, C$^{18}$O, and CS), we also computed results using the the (non-LTE) LVG method (Figure \ref{fig:discMaps}). For $^{13}$CO and C$^{18}$O, the Moment 0 maps (and HRO outcomes) were unaffected. However, for $^{12}$CO and CS there was some change. Particularly, the optical depth of the CS map increased, resulting in an increase in the PRS value (including a transition from negative to positive $Z_x$ in the top 10\% of pixels). Meanwhile, the $^{12}$CO emission (which was already optically thick in LTE) became somewhat less uniform. These results, though slightly different from the LTE case, are generally consistent with our main conclusion that the value of $Z_x$ is strongly linked to the optical depth of the observation. Future work is required for a more detailed assessment of how taking into account non-LTE effects generally affects the results for each tracer.
    \item The introduction of a Gaussian beam convolution (Figure \ref{fig:conv2}) to our synthetic maps tended to decrease the value of $Z_x$, though this effect is somewhat counterbalanced for larger (0.4 pc, 0.6 pc FWHM) beams due to washing out of some high density structures perpendicular to the mean magnetic field. Also of note, the sensitivity of an observation (as studied by testing intensity-based cuts at the top 50\%, 30\%, and 10\% of pixels) can have an important effect on the value of $Z_x$. 
    The effects of beam convolution and sensitivity limit need to be taken into account when using numerical simulations to interpret observations, such as those from \citet{fissel2019} for Vela C. 
\end{enumerate}

\section*{Acknowledgements}
RRM is supported by SOFIA grants (07-0235 and 09-0117) and an ALMA SOS award.
This work has been partially performed in support of developing analysis tools for the BLAST (Balloon-borne Large Aperture Sub- millimeter Telescope) collaboration. BLAST is supported by NASA under award numbers NNX13AE50G and 80NSSC18K0481.
RRM would like to additionally acknowledge support from the Virginia Space Grant Consortium (VSGC) Graduate Fellowship Award, and helpful correspondence with Ka Ho (Andy) Lam, Duo Xu, Ilse Cleeves, Crystal Brogan, and Bruce Wu. 
CYC, LNF, and ZYL are supported in part by NSF AST-1815784. 
ZYL acknowledges support from NASA 80NSSC20K0533. 
LMF acknowledges the support of the Natural Sciences and Engineering Research Council of Canada (NSERC) through Discovery Grant RGPIN/06266-2020, and funding through the Queen’s University Research Initiation Grant.
This work was performed under the auspices of the U.S. Department of Energy (DOE) by Lawrence Livermore National Laboratory under Contract DE-AC52-07NA27344 (CYC). LLNL-JRNL-843247-DRAFT. 
The simulations presented in this paper were performed on the Rivanna computer cluster at the University of Virginia.

\section*{Data Availability}
The 3D MHD and 2D synthetic observation data produced and analyzed for this work are available from the corresponding author, upon request.



\bibliographystyle{mnras}
\bibliography{main} 

\begin{thebibliography}{}
\makeatletter
\relax
\def\mn@urlcharsother{\let\do\@makeother \do\$\do\&\do\#\do\^\do\_\do\%\do\~}
\def\mn@doi{\begingroup\mn@urlcharsother \@ifnextchar [ {\mn@doi@}
  {\mn@doi@[]}}
\def\mn@doi@[#1]#2{\def\@tempa{#1}\ifx\@tempa\@empty \href
  {http://dx.doi.org/#2} {doi:#2}\else \href {http://dx.doi.org/#2} {#1}\fi
  \endgroup}
\def\mn@eprint#1#2{\mn@eprint@#1:#2::\@nil}
\def\mn@eprint@arXiv#1{\href {http://arxiv.org/abs/#1} {{\tt arXiv:#1}}}
\def\mn@eprint@dblp#1{\href {http://dblp.uni-trier.de/rec/bibtex/#1.xml}
  {dblp:#1}}
\def\mn@eprint@#1:#2:#3:#4\@nil{\def\@tempa {#1}\def\@tempb {#2}\def\@tempc
  {#3}\ifx \@tempc \@empty \let \@tempc \@tempb \let \@tempb \@tempa \fi \ifx
  \@tempb \@empty \def\@tempb {arXiv}\fi \@ifundefined
  {mn@eprint@\@tempb}{\@tempb:\@tempc}{\expandafter \expandafter \csname
  mn@eprint@\@tempb\endcsname \expandafter{\@tempc}}}

\bibitem[\protect\citeauthoryear{{Andr{\'e}}, {Di Francesco}, {Ward-Thompson},
  {Inutsuka}, {Pudritz}  \& {Pineda}}{{Andr{\'e}} et~al.}{2014}]{andre2014}
{Andr{\'e}} P.,  {Di Francesco} J.,  {Ward-Thompson} D.,  {Inutsuka} S.~I.,
  {Pudritz} R.~E.,   {Pineda} J.~E.,  2014, in {Beuther} H.,  {Klessen} R.~S.,
  {Dullemond} C.~P.,   {Henning} T.,  eds, Protostars and Planets VI. p.~27
  (\mn@eprint {arXiv} {1312.6232}),
  \mn@doi{10.2458/azu_uapress_9780816531240-ch002}

\bibitem[\protect\citeauthoryear{{Ballesteros-Paredes}, {Klessen}, {Mac Low}
  \& {Vazquez-Semadeni}}{{Ballesteros-Paredes} et~al.}{2007}]{ballesteros2007}
{Ballesteros-Paredes} J.,  {Klessen} R.~S.,  {Mac Low} M.~M.,
  {Vazquez-Semadeni} E.,  2007, in {Reipurth} B.,  {Jewitt} D.,   {Keil} K.,
  eds, Protostars and Planets V. p.~63 (\mn@eprint {arXiv} {astro-ph/0603357})

\bibitem[\protect\citeauthoryear{{Chen}, {King}  \& {Li}}{{Chen}
  et~al.}{2016}]{chen2016}
{Chen} C.-Y.,  {King} P.~K.,   {Li} Z.-Y.,  2016, \mn@doi [\apj]
  {10.3847/0004-637X/829/2/84}, \href
  {https://ui.adsabs.harvard.edu/abs/2016ApJ...829...84C} {829, 84}

\bibitem[\protect\citeauthoryear{{Chen}, {Li}, {King}  \& {Fissel}}{{Chen}
  et~al.}{2017}]{chen2017}
{Chen} C.-Y.,  {Li} Z.-Y.,  {King} P.~K.,   {Fissel} L.~M.,  2017, \mn@doi
  [\apj] {10.3847/1538-4357/aa898e}, \href
  {https://ui.adsabs.harvard.edu/abs/2017ApJ...847..140C} {847, 140}

\bibitem[\protect\citeauthoryear{{Ching}, {Li}, {Heiles}, {Li}, {Qian}, {Yue},
  {Tang}  \& {Jiao}}{{Ching} et~al.}{2022}]{ching2022}
{Ching} T.~C.,  {Li} D.,  {Heiles} C.,  {Li} Z.~Y.,  {Qian} L.,  {Yue} Y.~L.,
  {Tang} J.,   {Jiao} S.~H.,  2022, \mn@doi [\nat]
  {10.1038/s41586-021-04159-x}, \href
  {https://ui.adsabs.harvard.edu/abs/2022Natur.601...49C} {601, 49}

\bibitem[\protect\citeauthoryear{{Crutcher}}{{Crutcher}}{1999}]{crutcher1999}
{Crutcher} R.~M.,  1999, \mn@doi [\apj] {10.1086/307483}, \href
  {https://ui.adsabs.harvard.edu/abs/1999ApJ...520..706C} {520, 706}

\bibitem[\protect\citeauthoryear{{Crutcher}}{{Crutcher}}{2012}]{crutcher2012}
{Crutcher} R.~M.,  2012, \mn@doi [\araa] {10.1146/annurev-astro-081811-125514},
  \href {http://adsabs.harvard.edu/abs/2012ARA%26A..50...29C} {50, 29}

\bibitem[\protect\citeauthoryear{{Davis} \& {Greenstein}}{{Davis} \&
  {Greenstein}}{1951}]{davis1951}
{Davis} Leverett J.,  {Greenstein} J.~L.,  1951, \mn@doi [\apj]
  {10.1086/145464}, \href
  {https://ui.adsabs.harvard.edu/abs/1951ApJ...114..206D} {114, 206}

\bibitem[\protect\citeauthoryear{{Falgarone}, {Troland}, {Crutcher}  \&
  {Paubert}}{{Falgarone} et~al.}{2008}]{falgarone2008}
{Falgarone} E.,  {Troland} T.~H.,  {Crutcher} R.~M.,   {Paubert} G.,  2008,
  \mn@doi [\aap] {10.1051/0004-6361:200809577}, \href
  {https://ui.adsabs.harvard.edu/abs/2008A&A...487..247F} {487, 247}

\bibitem[\protect\citeauthoryear{{Field}}{{Field}}{1956}]{field1965}
{Field} G.~B.,  1956, \mn@doi [\apj] {10.1086/146261}, \href
  {https://ui.adsabs.harvard.edu/abs/1956ApJ...124..555F} {124, 555}

\bibitem[\protect\citeauthoryear{{Fissel} et~al.,}{{Fissel}
  et~al.}{2019}]{fissel2019}
{Fissel} L.~M.,  et~al., 2019, \mn@doi [\apj] {10.3847/1538-4357/ab1eb0}, \href
  {https://ui.adsabs.harvard.edu/abs/2019ApJ...878..110F} {878, 110}

\bibitem[\protect\citeauthoryear{{Fuente} et~al.,}{{Fuente}
  et~al.}{2019}]{fuente2019}
{Fuente} A.,  et~al., 2019, \mn@doi [\aap] {10.1051/0004-6361/201834654}, \href
  {https://ui.adsabs.harvard.edu/abs/2019A&A...624A.105F} {624, A105}

\bibitem[\protect\citeauthoryear{{Galitzki} et~al.,}{{Galitzki}
  et~al.}{2014}]{galitzki2014}
{Galitzki} N.,  et~al., 2014, in {Stepp} L.~M.,  {Gilmozzi} R.,   {Hall} H.~J.,
   eds,  Society of Photo-Optical Instrumentation Engineers (SPIE) Conference
  Series Vol. 9145, Ground-based and Airborne Telescopes V. p. 91450R
  (\mn@eprint {arXiv} {1407.3815}), \mn@doi{10.1117/12.2054759}

\bibitem[\protect\citeauthoryear{{Goodman}, {Crutcher}, {Heiles}, {Myers}  \&
  {Troland}}{{Goodman} et~al.}{1989}]{goodman1989}
{Goodman} A.~A.,  {Crutcher} R.~M.,  {Heiles} C.,  {Myers} P.~C.,   {Troland}
  T.~H.,  1989, \mn@doi [\apjl] {10.1086/185401}, \href
  {https://ui.adsabs.harvard.edu/abs/1989ApJ...338L..61G} {338, L61}

\bibitem[\protect\citeauthoryear{{Hartmann}, {Ballesteros-Paredes}  \&
  {Bergin}}{{Hartmann} et~al.}{2001}]{hartmann2001}
{Hartmann} L.,  {Ballesteros-Paredes} J.,   {Bergin} E.~A.,  2001, \mn@doi
  [\apj] {10.1086/323863}, \href
  {https://ui.adsabs.harvard.edu/abs/2001ApJ...562..852H} {562, 852}

\bibitem[\protect\citeauthoryear{{Heiles} \& {Crutcher}}{{Heiles} \&
  {Crutcher}}{2005}]{heiles2005}
{Heiles} C.,  {Crutcher} R.,  2005, in {Wielebinski} R.,  {Beck} R.,  eds, ,
  Vol.~664, Cosmic Magnetic Fields.
p.~137, \mn@doi{10.1007/3540313966_7}

\bibitem[\protect\citeauthoryear{{Heiles} \& {Troland}}{{Heiles} \&
  {Troland}}{2004}]{heiles2004}
{Heiles} C.,  {Troland} T.~H.,  2004, \mn@doi [\apjs] {10.1086/381753}, \href
  {https://ui.adsabs.harvard.edu/abs/2004ApJS..151..271H} {151, 271}

\bibitem[\protect\citeauthoryear{{Hennebelle} \& {P{\'e}rault}}{{Hennebelle} \&
  {P{\'e}rault}}{2000}]{hennebelle2000}
{Hennebelle} P.,  {P{\'e}rault} M.,  2000, \aap, \href
  {https://ui.adsabs.harvard.edu/abs/2000A&A...359.1124H} {359, 1124}

\bibitem[\protect\citeauthoryear{{Hoang} \& {Lazarian}}{{Hoang} \&
  {Lazarian}}{2009}]{hoang2009}
{Hoang} T.,  {Lazarian} A.,  2009, \mn@doi [\apj]
  {10.1088/0004-637X/697/2/1316}, \href
  {https://ui.adsabs.harvard.edu/abs/2009ApJ...697.1316H} {697, 1316}

\bibitem[\protect\citeauthoryear{{Hu}, {Yuen}, {Lazarian}, {Fissel}, {Jones}
  \& {Cunningham}}{{Hu} et~al.}{2019a}]{hu2019a}
{Hu} Y.,  {Yuen} K.~H.,  {Lazarian} A.,  {Fissel} L.~M.,  {Jones} P.~A.,
  {Cunningham} M.~R.,  2019a, \mn@doi [\apj] {10.3847/1538-4357/ab41f2}, \href
  {https://ui.adsabs.harvard.edu/abs/2019ApJ...884..137H} {884, 137}

\bibitem[\protect\citeauthoryear{{Hu}, {Yuen}  \& {Lazarian}}{{Hu}
  et~al.}{2019b}]{hu2019b}
{Hu} Y.,  {Yuen} K.~H.,   {Lazarian} A.,  2019b, \mn@doi [\apj]
  {10.3847/1538-4357/ab4b5e}, \href
  {https://ui.adsabs.harvard.edu/abs/2019ApJ...886...17H} {886, 17}

\bibitem[\protect\citeauthoryear{{Inoue}, {Inutsuka}  \& {Koyama}}{{Inoue}
  et~al.}{2007}]{inoue2007}
{Inoue} T.,  {Inutsuka} S.-i.,   {Koyama} H.,  2007, \mn@doi [\apjl]
  {10.1086/514816}, \href
  {https://ui.adsabs.harvard.edu/abs/2007ApJ...658L..99I} {658, L99}

\bibitem[\protect\citeauthoryear{{Jow}, {Hill}, {Scott}, {Soler}, {Martin},
  {Devlin}, {Fissel}  \& {Poidevin}}{{Jow} et~al.}{2018}]{jow2018}
{Jow} D.~L.,  {Hill} R.,  {Scott} D.,  {Soler} J.~D.,  {Martin} P.~G.,
  {Devlin} M.~J.,  {Fissel} L.~M.,   {Poidevin} F.,  2018, \mn@doi [\mnras]
  {10.1093/mnras/stx2736}, \href
  {https://ui.adsabs.harvard.edu/abs/2018MNRAS.474.1018J} {474, 1018}

\bibitem[\protect\citeauthoryear{{King}, {Fissel}, {Chen}  \& {Li}}{{King}
  et~al.}{2018}]{king2018}
{King} P.~K.,  {Fissel} L.~M.,  {Chen} C.-Y.,   {Li} Z.-Y.,  2018, \mn@doi
  [\mnras] {10.1093/mnras/stx3096}, \href
  {https://ui.adsabs.harvard.edu/abs/2018MNRAS.474.5122K} {474, 5122}

\bibitem[\protect\citeauthoryear{{K{\"o}rtgen} \& {Banerjee}}{{K{\"o}rtgen} \&
  {Banerjee}}{2015}]{kortgen2015}
{K{\"o}rtgen} B.,  {Banerjee} R.,  2015, \mn@doi [\mnras]
  {10.1093/mnras/stv1200}, \href
  {https://ui.adsabs.harvard.edu/abs/2015MNRAS.451.3340K} {451, 3340}

\bibitem[\protect\citeauthoryear{{Lazarian} \& {Hoang}}{{Lazarian} \&
  {Hoang}}{2007}]{lazarian2007}
{Lazarian} A.,  {Hoang} T.,  2007, \mn@doi [\mnras]
  {10.1111/j.1365-2966.2007.11817.x}, \href
  {http://adsabs.harvard.edu/abs/2007MNRAS.378..910L} {378, 910}

\bibitem[\protect\citeauthoryear{{Lazarian} \& {Yuen}}{{Lazarian} \&
  {Yuen}}{2018}]{lazarian2018}
{Lazarian} A.,  {Yuen} K.~H.,  2018, \mn@doi [\apj] {10.3847/1538-4357/aaa241},
  \href {https://ui.adsabs.harvard.edu/abs/2018ApJ...853...96L} {853, 96}

\bibitem[\protect\citeauthoryear{{Mac Low} \& {Klessen}}{{Mac Low} \&
  {Klessen}}{2004}]{maclow2004}
{Mac Low} M.-M.,  {Klessen} R.~S.,  2004, \mn@doi [Reviews of Modern Physics]
  {10.1103/RevModPhys.76.125}, \href
  {https://ui.adsabs.harvard.edu/abs/2004RvMP...76..125M} {76, 125}

\bibitem[\protect\citeauthoryear{{Maret}, {Bergin}  \& {Lada}}{{Maret}
  et~al.}{2006}]{maret2006}
{Maret} S.,  {Bergin} E.~A.,   {Lada} C.~J.,  2006, \mn@doi [\nat]
  {10.1038/nature04919}, \href
  {https://ui.adsabs.harvard.edu/abs/2006Natur.442..425M} {442, 425}

\bibitem[\protect\citeauthoryear{{McKee} \& {Ostriker}}{{McKee} \&
  {Ostriker}}{2007}]{mckee2007}
{McKee} C.~F.,  {Ostriker} E.~C.,  2007, \mn@doi [\araa]
  {10.1146/annurev.astro.45.051806.110602}, \href
  {https://ui.adsabs.harvard.edu/abs/2007ARA&A..45..565M} {45, 565}

\bibitem[\protect\citeauthoryear{{Mestel} \& {Spitzer}}{{Mestel} \&
  {Spitzer}}{1956}]{mestel1956}
{Mestel} L.,  {Spitzer} L. J.,  1956, \mn@doi [\mnras]
  {10.1093/mnras/116.5.503}, \href
  {https://ui.adsabs.harvard.edu/abs/1956MNRAS.116..503M} {116, 503}

\bibitem[\protect\citeauthoryear{{Morgan}, {Moore}, {Allsopp}  \&
  {Eden}}{{Morgan} et~al.}{2013}]{morgan2013}
{Morgan} L.~K.,  {Moore} T.~J.~T.,  {Allsopp} J.,   {Eden} D.~J.,  2013,
  \mn@doi [\mnras] {10.1093/mnras/sts098}, \href
  {https://ui.adsabs.harvard.edu/abs/2013MNRAS.428.1160M} {428, 1160}

\bibitem[\protect\citeauthoryear{{Mouschovias} \& {Spitzer}}{{Mouschovias} \&
  {Spitzer}}{1976}]{mouschovias1976}
{Mouschovias} T.~C.,  {Spitzer} L. J.,  1976, \mn@doi [\apj] {10.1086/154835},
  \href {https://ui.adsabs.harvard.edu/abs/1976ApJ...210..326M} {210, 326}

\bibitem[\protect\citeauthoryear{{Nakamura} \& {Li}}{{Nakamura} \&
  {Li}}{2008}]{nakamura2008}
{Nakamura} F.,  {Li} Z.-Y.,  2008, \mn@doi [\apj] {10.1086/591641}, \href
  {https://ui.adsabs.harvard.edu/abs/2008ApJ...687..354N} {687, 354}

\bibitem[\protect\citeauthoryear{{Ostriker}, {Stone}  \& {Gammie}}{{Ostriker}
  et~al.}{2001}]{ostriker2001}
{Ostriker} E.~C.,  {Stone} J.~M.,   {Gammie} C.~F.,  2001, \mn@doi [\apj]
  {10.1086/318290}, \href
  {https://ui.adsabs.harvard.edu/abs/2001ApJ...546..980O} {546, 980}

\bibitem[\protect\citeauthoryear{{Planck Collaboration Int. XIX}}{{Planck
  Collaboration Int. XIX}}{2015}]{planckXIX}
{Planck Collaboration Int. XIX} 2015, \mn@doi [A\&A]
  {10.1051/0004-6361/201424082}, 576, A104

\bibitem[\protect\citeauthoryear{{Planck Collaboration Int. XXXV}}{{Planck
  Collaboration Int. XXXV}}{2016}]{planckXXXV}
{Planck Collaboration Int. XXXV} 2016, \mn@doi [A\&A]
  {10.1051/0004-6361/201525896}, 586, A138

\bibitem[\protect\citeauthoryear{{Planck Collaboration} et~al.,}{{Planck
  Collaboration} et~al.}{2015}]{planckXX}
{Planck Collaboration} et~al., 2015, \mn@doi [\aap]
  {10.1051/0004-6361/201424086}, \href
  {https://ui.adsabs.harvard.edu/abs/2015A&A...576A.105P} {576, A105}

\bibitem[\protect\citeauthoryear{{Scalo}}{{Scalo}}{1985}]{scalo1985}
{Scalo} J.~M.,  1985, in {Black} D.~C.,  {Matthews} M.~S.,  eds, Protostars and
  Planets II. pp 201--296

\bibitem[\protect\citeauthoryear{{Sch{\"o}ier}, {van der Tak}, {van Dishoeck}
  \& {Black}}{{Sch{\"o}ier} et~al.}{2005}]{schoier2005}
{Sch{\"o}ier} F.~L.,  {van der Tak} F.~F.~S.,  {van Dishoeck} E.~F.,   {Black}
  J.~H.,  2005, \mn@doi [\aap] {10.1051/0004-6361:20041729}, \href
  {https://ui.adsabs.harvard.edu/abs/2005A&A...432..369S} {432, 369}

\bibitem[\protect\citeauthoryear{{Shu}, {Adams}  \& {Lizano}}{{Shu}
  et~al.}{1987}]{shu1987}
{Shu} F.~H.,  {Adams} F.~C.,   {Lizano} S.,  1987, \mn@doi [\araa]
  {10.1146/annurev.aa.25.090187.000323}, \href
  {https://ui.adsabs.harvard.edu/abs/1987ARA&A..25...23S} {25, 23}

\bibitem[\protect\citeauthoryear{{Soler} \& {Hennebelle}}{{Soler} \&
  {Hennebelle}}{2017}]{soler2017b}
{Soler} J.~D.,  {Hennebelle} P.,  2017, \mn@doi [\aap]
  {10.1051/0004-6361/201731049}, \href
  {https://ui.adsabs.harvard.edu/abs/2017A&A...607A...2S} {607, A2}

\bibitem[\protect\citeauthoryear{{Soler}, {Hennebelle}, {Martin},
  {Miville-Desch{\^e}nes}, {Netterfield}  \& {Fissel}}{{Soler}
  et~al.}{2013}]{soler2013}
{Soler} J.~D.,  {Hennebelle} P.,  {Martin} P.~G.,  {Miville-Desch{\^e}nes}
  M.~A.,  {Netterfield} C.~B.,   {Fissel} L.~M.,  2013, \mn@doi [\apj]
  {10.1088/0004-637X/774/2/128}, \href
  {https://ui.adsabs.harvard.edu/abs/2013ApJ...774..128S} {774, 128}

\bibitem[\protect\citeauthoryear{{Soler} et~al.,}{{Soler}
  et~al.}{2017}]{soler2017}
{Soler} J.~D.,  et~al., 2017, \mn@doi [\aap] {10.1051/0004-6361/201730608},
  \href {https://ui.adsabs.harvard.edu/abs/2017A&A...603A..64S} {603, A64}

\bibitem[\protect\citeauthoryear{{Stone}, {Gardiner}, {Teuben}, {Hawley}  \&
  {Simon}}{{Stone} et~al.}{2008}]{stone2008}
{Stone} J.~M.,  {Gardiner} T.~A.,  {Teuben} P.,  {Hawley} J.~F.,   {Simon}
  J.~B.,  2008, \mn@doi [\apjs] {10.1086/588755}, \href
  {https://ui.adsabs.harvard.edu/abs/2008ApJS..178..137S} {178, 137}

\bibitem[\protect\citeauthoryear{{Troland} \& {Crutcher}}{{Troland} \&
  {Crutcher}}{2008}]{troland2008}
{Troland} T.~H.,  {Crutcher} R.~M.,  2008, \mn@doi [\apj] {10.1086/587546},
  \href {https://ui.adsabs.harvard.edu/abs/2008ApJ...680..457T} {680, 457}

\bibitem[\protect\citeauthoryear{{Xu}, {Ji}  \& {Lazarian}}{{Xu}
  et~al.}{2019}]{xu2019}
{Xu} S.,  {Ji} S.,   {Lazarian} A.,  2019, \mn@doi [\apj]
  {10.3847/1538-4357/ab21be}, \href
  {https://ui.adsabs.harvard.edu/abs/2019ApJ...878..157X} {878, 157}

\makeatother
\end{thebibliography}




\appendix



\bsp	
\label{lastpage}
\end{document}